\documentclass{jaa}
\usepackage[square]{natbib}
\usepackage[colorlinks]{hyperref}
\usepackage{xcolor, wasysym, ulem}
\definecolor{LightCyan}{rgb}{0.88,1,1}
\definecolor{skyblue}{RGB}{0, 170, 255}
\definecolor{BlueViolet}{RGB}{138, 43,226}
\definecolor{MidnightBlue}{RGB}{0, 120, 255}
\definecolor{orange}{RGB}{255,140,0}
\definecolor{forestgreen}{RGB}{20,160,120}
\definecolor{gray50}{RGB}{127,127,127}
\usepackage{graphicx, graphics, psfrag}
\usepackage{lastpage}

\graphicspath{{figs/}}

\def\HI{{\sc Hi}}
\def\HII{{\sc Hii}}
\def\xb{\bar{x}_{\rm {HI}}}

\def\cl{\mathcal{C}_{\mathcal{\ell}}}
\def\l{\mathcal{\ell}}

\def\V{\mathcal{V}}
\def\u{\vec{U}}
\def\lsim{~\rlap{$<$}{\lower 1.0ex\hbox{$\sim$}}}

\def\gsim{~\rlap{$>$}{\lower 1.0ex\hbox{$\sim$}}}

\newcommand{\nhat}{{\hat{n}}}
\newcommand{\kvec}{\vec{k}}
\newcommand{\xvec}{\vec{x}}
\newcommand{\xhb}{\bar{x}_{\rm HI}}

\def\cl{{\mathit C}_{\ell}}

\def\n{\hat{\mathit{n}}}

\begin{document}\sloppy

\title{Probing early universe through redshifted 21-cm signal: Modelling and observational challenges}


\author{Abinash Kumar Shaw\textsuperscript{1,2,*}, Arnab Chakraborty\textsuperscript{3,*}, Mohd Kamran\textsuperscript{4}, Raghunath Ghara\textsuperscript{2},\\ Samir Choudhuri\textsuperscript{5,6}, Sk. Saiyad Ali\textsuperscript{7}, Srijita Pal\textsuperscript{1}, Abhik Ghosh\textsuperscript{8}, Jais Kumar\textsuperscript{9,10}, Prasun Dutta\textsuperscript{9},\\ Anjan Kumar Sarkar\textsuperscript{11,12}}
\affilOne{\textsuperscript{1}Department of Physics, Indian Institute of Technology Kharagpur, Kharagpur  721302, India.\\}
\affilTwo{\textsuperscript{2}Astrophysics Research Centre, Open University of Israel, Ra'anana 4353701, Israel.\\}
\affilThree{\textsuperscript{3}Department of Physics and McGill Space Institute, McGill University, Montreal, QC, Canada H3A 2T8.\\}
\affilFour{\textsuperscript{4}Department of Astronomy, Astrophysics \& Space Engineering, Indian Institute of Technology Indore, Indore 453552, India.\\}
\affilFive{\textsuperscript{5}School of Physics and Astronomy, Queen Mary University of London, London E1 4NS, U.K.\\}
\affilSix{\textsuperscript{6}Department of Physics, Indian Institute of Technology Madras, Chennai 600036, India.\\}
\affilSeven{\textsuperscript{7}Department of Physics, Jadavpur University, Kolkata 700032, India.\\}
\affilEight{\textsuperscript{8}Department of Physics, Banwarilal Bhalotia College, Asansol, West Bengal, India.\\}
\affilNine{\textsuperscript{9} Department of Physics, Indian Institute of Technology (Banaras Hindu University), Varanasi - 221005, India.\\}
\affilTen{\textsuperscript{10}Department of Physics, K.N. Govt. P.G. College, Gyanpur, Bhadohi - 221304, India.\\}
\affilEleven{\textsuperscript{11}Raman Research Institute, Sadashivnagar, Bangalore, Karnataka 560080, India.\\}
\affilTwelve{\textsuperscript{12}National Centre for Radio Astrophysics, TIFR, Pune University Campus, Post Bag 3, Pune 411 007, India.\\}


\twocolumn[{

\maketitle

\corres{\href{mailto:abinashkumarshaw@gmail.com}{abinashkumarshaw@gmail.com}} \corres{\href{mailto:arnab.phy.personal@gmail.com}{arnab.phy.personal@gmail.com}}
\msinfo{DD MM YYYY}{DD MM YYYY}

\begin{abstract}
Cosmic Dawn (CD) and Epoch of Reionization (EoR) are the most important part of the cosmic history during which the first luminous structures emerged. These first objects heated and ionized the neutral atomic hydrogen in the intergalactic medium. The redshifted 21-cm radiation from the atomic hydrogen provides an excellent direct probe to study the evolution of neutral hydrogen (\HI) and thus reveal the nature of the first luminous objects, their evolution and role in this last phase transition of the universe and formation and evolution of the structures thereafter. Direct mapping of the \HI~density during the CD-EoR is rather difficult  with the current and forthcoming instruments due to stronger foreground and other observational contamination. The first detection of this redshifted \HI~signal is expected to be done through statistical estimators. Given the utmost importance of the detection and analysis of the redshifted 21-cm signal, physics of CD-EoR is considered as one of the objectives of upcoming the SKA-Low telescope. This paper summarizes the collective effort of Indian Astronomers to understand the origin of the redshifted 21-cm signal, sources of first ionizing photons, their propagation through the IGM, various cosmological effects on the expected 21-cm signal, various statistical measures of the signal like power spectrum, bispectrum, etc. A collective effort on detection of such signal by developing estimators of the statistical measures with rigorous assessment of their expected uncertainties, various challenges like that of the large foreground emission and calibration issues are also discussed. Various versions of the detection methods discussed here have also been used in practice with the Giant Meterwave Radio Telescope with successful assessment of the foreground contamination and upper limits on the matter density in reionization and post-reionization era. The collective efforts compiled here has been a large part of the global effort to prepare proper observational technique and analysis procedure for the first light of the CD-EoR through the SKA-Low.
\end{abstract}


}]



\doinum{12.3456/s78910-011-012-3}
\artcitid{\#\#\#\#}
\volnum{000}
\year{0000}
\pgrange{1--\pageref{LastPage}}
\setcounter{page}{1}
\lp{\pageref{LastPage}}


\section{Introduction}\label{sec:intro}

In the standard model of cosmology, the universe started with a hot and dense state. As the universe expands adiabatically, it cooled down to a temperature at which the matter components of the universe (electrons and protons) combined to form neutral atomic hydrogen (\HI). This occurred at redshift $z\approx 1100$ \citep[e.g.][]{peebles_1993} when radiations got decoupled from matter and travelled freely across the inter-galactic medium (IGM). Afterglow of the surface of last scattering is observed as the Cosmic Microwave Background (CMB). The CMB observations indicates that tiny fluctuations ($\sim 10^{-6}$) were present in the matter density field during matter-radiation decoupling. Afterwards, the universe remained completely dark and neutral for a long time till these fluctuations have grown sufficiently to form first luminous objects which lit up the universe at $z\approx 20$ \citep[see e.g.][]{ghara15a, ghara15b}. This marks the `Cosmic Dawn' (CD) during which the X-rays from the first objects (stars, quasars, galaxies etc.) heated up the inter-galactic medium (IGM). Sooner, the ionizing UV photons from these sources also started leaking into the IGM and gradually ionized all the neutral hydrogen (\HI) in the IGM. This last phase change in the ionization state of the universe is coined as the `Epoch of Reionization' (EoR). Afterwards, the universe remained ionized on cosmological scales as we see it today. This last stage is also known as `Post-reionization epoch'. CMB observations give us a clearer picture of how the universe was in it's primary stages and observations like the galaxy-redshift surveys show the present state of the universe. However we know only a little about the CD-EoR. 

Our present knowledge of CD-EoR is guided by a few indirect observations such as measurements of Thomson scattering optical depth from CMB observations \citep[e.g.][]{douspis2015,planck2016,planck2018}, observation of Gunn-Peterson troughs in the high-$z$ quasars \citep[e.g.][]{becker2001,fan2002,fan2006} and measurements of luminosity functions from the high-$z$ Ly-$\alpha$ emitters \citep[e.g.][]{hu2002, malhotra2004, ota2007, ouchi2010, choudhury2015}. These observations commonly suggest that EoR has started roughly at $z\approx 13$ and ended by $z\approx 6$ \citep[e.g.][]{robertson2013,robertson2015,Mondal2015,mitra2018a,mitra2018b}. However a deeper insight of these epochs are required in order to understand these primary stages of the structure formation. \HI~being the most abundant element in the IGM during these epochs, the 21-cm radiation which originates due to hyperfine transition in \HI~proves to be most promising probe of CD-EoR. The CD-EoR 21-cm signal is highly sensitive to the properties of the first astrophysical sources and their interactions with the IGM. Therefore mapping out the intensity distribution of the redshifted 21-cm radiation from \HI~in the IGM provides a unique and direct way to study CD-EoR \citep[e.g.][]{sunyaev1975, hogan1979, scott1990, bharadwaj2001, bharadwaj2001b}. Given such important information that the observations of redshifted 21-cm signal is expected to reveal, it's in one of the major science goals of the upcoming Square Kilometer Array (SKA-Low) \citep{koopmans2015}. It is expected that the direct observations of the CD-EoR 21-cm signal would be able to answer various fundamental questions related to the progress of the heating and reionization of the IGM, properties of sources involved and their evolution etc.

A substantial effort is ongoing with the SKA pathfinder radio-interferometers such as uGMRT\citep{Swarup_1991, Yashwant_2017}, PAPER\citep{parsons2010}, LOFAR\citep{vanhaarlem2013}, MWA\citep{tingay2013}, HERA\citep{deboer2017} and NenuFAR\citep{nenufar} to detect the CD-EoR 21-cm signal. However most of them are not suited for the CD observations due to their limited sensitivity and range of operation. The upcoming SKA-Low will be a giant leap in terms of sensitivity as well as it will have a large frequency bandwidth ($50-350 \, {\rm MHz}$) to cover both CD and EoR. Unfortunately, even with first phase of SKA-Low, a direct detection of the signal is not possible due to $\sim 10^4-10^5$ times stronger galactic and extra-galactic foregrounds \citep[e.g.][]{Ali_2008, Bern09, Ghosh_2012}. Therefore the current experiments aim to observe the signal by measuring its statistics, majorly the power spectrum (PS) \citep[e.g.][]{bharadwaj2001b, Bharadwaj_2004, Bharadwaj_2005}. However, only few weak upper limits on the PS amplitudes have been reported to date (e.g. GMRT: \citealt{paciga2011, paciga2013}; LOFAR: \citealt{yatwatta2013, patil2017, gehlot2019, Mertens2020} MWA: \citealt{li2019, barry2019, Trott_2020}; PAPER: \citealt{cheng2018, kolopanis2019}; HERA: \citealt{HERA2021}). In its first phase, SKA-Low shall be able to measure the 21-cm power spectrum with high precision at different redshifts within relatively less observation time. One would additionally compute higher-order statistics such as bispectrum \citep[e.g.][]{majumdar18, majumdar20, kamran21, kamran22, 2019JCAP...02..058G}, trispectrum \citep{Mondal2016, Mondal2017} etc. which can provide additional information of these epochs. Moreover, the second phase of the SKA-Low is expected to produce images of the 21-cm signal maps from CD and EoR. One can then use various image-based statistical tools such as Minkowski functional \citep[e.g.][]{kapathia2018, kapathia2019, kapathia2021} and Largest Cluster Statistics \citep[e.g.][]{bag2018, bag2019, pathak2022} etc. to extract maximum information out of the signal.

The \HI~intensity mapping of the post-reionization $(z \le 6)$ 21-cm signal holds the potential to probe the large-scale structures and constrain various cosmological parameters \citep{bharadwaj2001, bharadwaj2001b,loeb08,Bh09, Visbal_2009,Vill15}. It can independently probe the expansion history of the Universe by measuring the Baryon Acoustic Oscillation (BAO) in the 21-cm power spectrum (PS) \citep{w08, Chang08, Seo_2010}.

Several efforts have already been carried out towards detecting the post-reionization 21-cm signal. \citet{pen09} first detected the signal by cross-correlating the \HI~Parkes All Sky Survey (HIPASS) and the Six degree Field Galaxy Redshift Survey (6dFGRS; \citealt{jones2004}). At a higher redshift $(z \sim 0.8)$, the detection of the cross power spectrum has been presented in \citep{Chang08} using 21-cm intensity maps acquired at the Green Bank Telescope (GBT) and the DEEP2 galaxy survey. Further improvement on these measurements was carried out \citep{masui2013} by cross-correlating new intensity mapping data from the WiggleZ Dark Energy Survey \citep{drinkwater2010}. The auto-power spectrum measurement of 21-cm intensity fluctuation maps acquired with GBT has been used to constrain neutral hydrogen fluctuations at $z \sim 0.8$ \citep{SW13}. These measurements were conducted using single-dish telescopes in the low-redshift $(z < 1)$ regime, where we already have optical surveys.  Next-generation intensity mapping surveys with SKA-Mid \citep{SKA15} will have the potential to open up a large cosmological window at the post-reionization epoch, allowing us to detect the 21-cm signal with a high level of accuracy.

Rest of the article is arranged in the following way. We start section \ref{sec:models} discussing analytical and numerical models of the 21-cm signal. Next we present different statistical tools to quantify the signal in section \ref{sec:stats}. We mention the observational challenges in detection of the \HI~21-cm signal in section \ref{sec:detection}. We briefly mention the different foreground contribution to the observed signal. In section \ref{sec:upper_lim}, we quote the present upper limits on the signal with the current telescopes and also discuss the prospects of measuring signal statistics using SKA-Low. We finally summarize the document in section \ref{sec:summ}. 


\section{Modelling 21-cm signal}\label{sec:models}

The redshifted 21-cm radiation acts as a proxy to the \HI~distribution in the IGM which almost follows the underlying matter density field during CD. However during EoR, the \HI~distribution is largely determined by the ionized regions.
The \HI~21-cm signal is quantified using the differential brightness temperature \citep{rybicki1979} observed against CMB at a frequency $\nu$ and along a direction $\nhat$. This can be written as
\begin{eqnarray}
    T_b(\nhat r_{\nu}, \nu)&= &  \bar{T}(z)  \xhb (z)[1+\delta_{\rm HI}(\nhat r_{\nu}, z)] \left( 1- \frac{T_{\gamma}}{T_{\rm s}} \right) \,\nonumber \\  &\times & \left[1-\frac{1+z}{H(z)}\frac{\partial v_{\parallel}(\nhat r_{\nu}, z)}{\partial r_{\nu}}\right]\,,
\label{eq:bt}
\end{eqnarray}
where $\nu$ is the frequency of observation and related to the redshift $z$ as $\nu=1420/(1+z)$ {\rm MHz}. $r_{\nu}$ is the comoving distance to the redshift $z$, $\xhb$ is the mean neutral hydrogen fraction, $\delta_{\rm HI}$ is the \HI~density contrast, $H(z)$ is the Hubble parameter, $\frac{\partial v_{\parallel}}{\partial r_{\nu}}$ is the gradient of peculiar velocity along the line-of-sight (LoS) direction and the characteristic \HI~brightness temperature
\begin{equation}
 \bar{T}(z) = 4.0\,\text{mK} \, (1+z)^2 \, \left( \frac{1-Y_P}{0.75}
\right) \, \left( \frac{\Omega_b h^2}{0.020} \right) \left(
\frac{0.7}{h} \right) \left( \frac{H_0}{H(z)} \right)\,,
\label{eq:ct}
\end{equation}
where the symbols have their usual meaning. Note that $\bar{T}(z)$ depends only on the background cosmological model. The CD-EoR 21-cm signal also depends on the factor $(1-T_{\gamma}/T_{\rm s})$ in eq. (\ref{eq:bt}) where $T_{\gamma}=T_{\gamma 0} (1+z)$ is the CMB temperature and $T_{\rm s}$ is the spin temperature of the \HI~gas in the IGM. The spin temperature is not a thermodynamical quantity, rather it is defined by the relative population of \HI~atoms between two hyperfine states \textit{i.e.}
\begin{equation}
    \frac{n_1}{n_0} = \frac{g_1}{g_0} \exp{(-T_*/T_{\rm s})}~.
    \label{eq:spin_temp}
\end{equation}
Here $g_1=3$ and $g_0=1$ are the degeneracy factors of the excited states and ground states with $n_1$ and $n_0$ being numbers of \HI~atoms in the respective states. $T_* = 0.068\, {\rm K} = h_{\rm p} \nu_e/k_{\rm B}$, where $h_{\rm P}$ is
the Planck constant, $\nu_e =1420 ~{\rm MHz}$ and $k_{\rm B}$ is the Boltzmann constant. There are three major physical processes which controls the level population of \HI~atoms and couples $T_{\rm s}$ to either the gas kinetic temperature $T_{\rm K}$ or $T_{\gamma}$ or the Ly-$\alpha$ color temperature $T_{\alpha}$ during various stages. Computation shows that the $T_{\rm s}$ depends on the other three temperatures as \citep[e.g.][]{pritchard2012}
\begin{equation}
    T_{\rm s}^{-1} = \frac{T_{\gamma}^{-1} + x_c T_{\rm K}^{-1} + x_{\alpha} T_{\alpha}^{-1}}{1 + x_c + x_{\alpha}}~,
    \label{eq:Ts}
\end{equation}
where $x_c$ and $x_{\alpha}$ denotes the strength of the different processes. During the initial stages of CD the Ly-$\alpha$ radiations from the sources electronically excited and de-excited the \HI~atoms due to resonant scattering. This redistributes the atoms between the two hyperfine levels and therefore couples the spin temperature to $T_{\alpha} \approx T_{\rm K}$. This is known as Wouthuysen-Field effect \citep{wouthuysen1952, field1958, chen2004} makes $x_{\alpha}$ dominant over all other coupling terms. Later at $z \approx 17$, the X-rays have been started heating IGM and $T_{\rm s}$ starts rising towards $T_{\gamma}$. Therefore the CD 21-cm signal is observed in absorption, however reionization starts as soon as $T_{\rm s}$ overshoots $T_{\gamma}$ and the EoR 21-cm signal is observed in emission. The evolution of the simulated mean brightness temperature $\bar{T}_b(z)$ during CD-EoR is shown in the bottom panel of the Figure \ref{fig:GRIZZLY_LC}. There are several experiments which aims to measure this global signal from CD-EoR. However the radio-interferometric instruments, such as SKA, are expected to observe the spatial fluctuations in the brightness temperature of the CD-EoR signal which is defined as 
\begin{equation}
\delta T_b(\nhat r_{\nu}, \nu)= T_b(\nhat r_{\nu}, \nu)- \bar{T}_b(\nu)~,
\label{eq:dTb}
\end{equation}
where $\bar{T}_b(\nu)\equiv \bar{T}_b(z)$, as shown in bottom panel of Figure \ref{fig:GRIZZLY_LC}.

The formation and evolution of the first sources (as mentioned in section \ref{sec:intro}) is expected to actively control the topology of the fluctuating CD-EoR signal $\delta T_b(\nhat r_{\nu}, \nu)$ \citep{barkana2000,furlanetto06,pritchard2012}. It is widely accepted that the star-forming galaxies during the CD-EoR are the primary source of UV photon production. Having a small mean free path, they can not photoionize \HI~much far into the IGM. Some of these galaxies are likely to have accreting supermassive or intermediate-mass black holes at their centers, and they will act as mini-quasars (mini-QSOs). The mini-QSOs can majorly produce high-energy X-ray photons, which have larger mean free path and heat up the IGM. Besides, the galaxies with a high star formation rate are likely to host binary systems such as the high-mass X-ray binaries (HMXBs), which also produce a copious amount of X-ray photons and interact with the IGM differently as compared to the mini-QSOs. 

The relative abundances of the different sources and their radiation processes can have distinct imprints on the evolution of \HI~in the IGM during CD-EoR. It is widely understood that the UV photons ionize \HI~efficiently than the X-ray photons. Hence, for example, if the universe is dominated by the UV radiations from the galaxies, it would become an so-called `inside-out' reionization scenario. This is because the UV photons first ionize the dense regions around the sources and then the ionization fronts advance further into the IGM. On the other hand, if the sources such as HMXBs producing hard X-ray photons with large mean free paths would have been most abundant in the early universe, they could cause the so-called `outside-in' reionization scenario \citep{Choudhury2009}. In this scenario, high energy X-ray photons would easily escape from their host galaxies and travel the large distance in the IGM to be redshifted to the lower frequency corresponding to the ionizing photons. Hence the under-dense regions ionize first, and as time goes on, the reionization slowly progresses into the denser regions. In practice, both of these scenarios contribute to the photo-ionization of IGM gas. However, the inside-out scenario driven by UV photons from the galaxies dominates the photo-ionization. On contrary, the X-ray photons contributes majorly to heat up the IGM thereby modulating the $T_{\rm s}$ of the \HI.

We now briefly summarize analytical and numerical methods to simulate $\delta T_b(\nhat r_{\nu}, \nu)$ of the CD-EoR 21-cm signal in the following subsections.

\subsection{Analytical Modelling}\label{subsec:Analytic}

According to eqs. (\ref{eq:bt}) and (\ref{eq:dTb}), one needs a perfect knowledge of $\bar{x}_{\rm HI}(z)$, $\delta_{\rm HI}(\nhat r_{\nu}, z)$, $T_{\rm s}(\nhat r_{\nu}, z)$ and $\partial v_{\parallel}(\nhat r_{\nu}, z)/\partial r_{\nu}$ to accurately model $\delta T_b(\nhat r_{\nu}, \nu)$. Due to complicated interplay between the cosmological and astrophysical processes, it is not straightforward to write analytical expressions for all of these quantities and hence for $\delta T_b$ during CD-EoR. However modelling of the statistics of the signal is possible under some simplified assumption. Here we briefly discuss a simplistic but effective model of the EoR 21-cm power spectrum as described in \citet{Bharadwaj_2005}. Note that power spectrum is the Fourier transform of the two-point correlation of a fluctuating field. We refer the reader to section \ref{subsec:2.3.1} below for a complete description of the power spectrum statistics.

\begin{figure}[ht!]
\centering
\includegraphics[width=0.5\textwidth, angle=0]{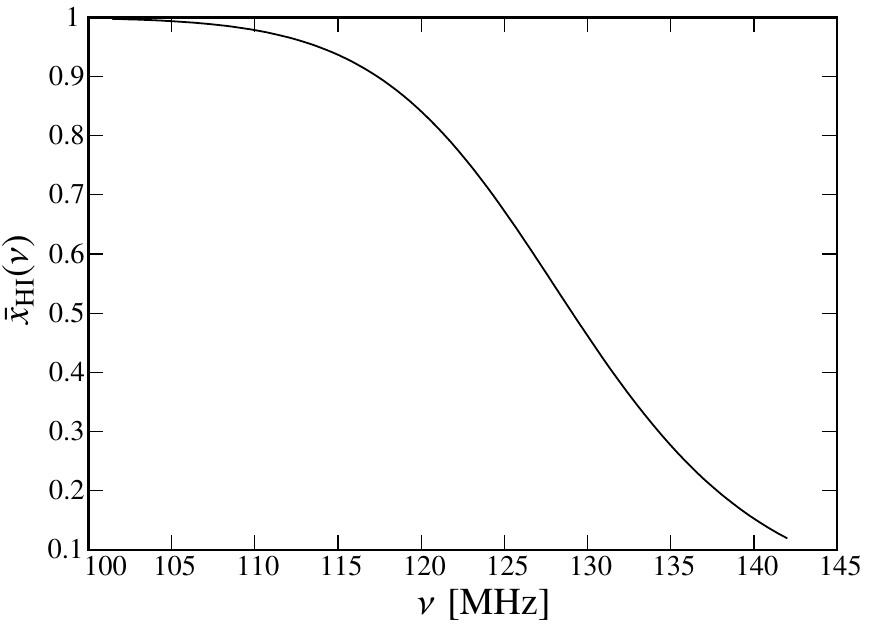}
\caption{This shows the evolution of the mean neutral fraction $\bar{x}_{\rm HI}$ with the frequency of the redshifted HI emission for the reionization model discussed in the text.}
\label{fig:nf}
\end{figure}

\begin{figure}[ht!]
\centering
\includegraphics[width=0.5\textwidth, angle=0]{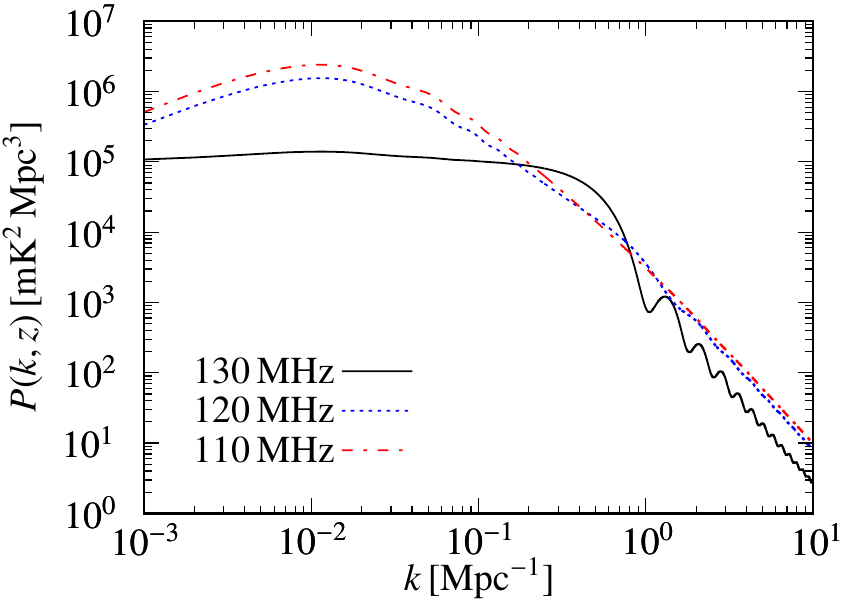}
\caption{Shows the expected \HI~power spectrum $P(k,z)$ from the epoch of reionization at different frequencies (as shown in the figure) for models of the \HI~distribution considered here. Here the $\mu$ dependence has been incorporated by using the average
value $\langle \mu^2 \rangle = 1/3$.}
\label{fig:eorpk}
\end{figure}

As mentioned before, assuming that the \HI~gas is heated well before it is reionized, and that the spin temperature is coupled to the gas temperature with $T_{\rm s} \gg T_{\gamma}$ CMB temperature so that  $(1-T_{\gamma}/T_s) \rightarrow 1$ (eq. \ref{eq:bt}). Now, the EoR 21-cm emission signal depends only on the \HI~density contrast and peculiar velocity. We assume that the hydrogen density and peculiar velocities follow the dark matter with possible bias $b_{\rm HI}$. Further, it is assumed that non-overlapping spheres of comoving radius $R$ are completely ionized, the centres of the spheres being clustered with a bias $b_c \geq 1$ relative to underlying dark matter distribution. One would expect the centres of the ionized spheres to be clustered, given the fact that one identify them with the locations of the first luminous objects which are believed to have formed at the peaks of the density fluctuations. Similar kind of models have used in the context of the effect of patchy reionization on the CMB \citep{knox1998,gruz1998} and \HI~emission in high redshifts \citep{Zaldarriaga_2004}.

In this simple reionization model, the ionization history is considered as $\bar{x}_{\rm HI}(z)={1}/{1+\exp((z-z_0)/\Delta z)}$ where $z_0=10$ and $\Delta z=0.5$ so that $50 \%$ of the hydrogen is reionized at a redshift $z=10$. The fraction of volume ionized $f_V$, the mean neutral fraction $\bar{x}_{\rm HI}$ and mean comoving number density of ionized spheres $\bar{n}_{\rm HI}$ are related as $f_V=1-\bar{x}_{\rm HI}=(4  \pi  R^3/3) \bar{n}_{\rm HI}$. The comoving size of the spheres $R$ increases with $z$ so that the mean comoving number density of the  ionized spheres remains constant. Using various assumptions mentioned above, the \HI~power spectrum of brightness temperature fluctuations have been modelled as \citep[for more details see][]{Bharadwaj_2005}

\begin{equation}
\begin{split}
    P(\kvec,z)=& \bar{T}^2(z)  \left[b_{\rm HI}\,\bar{x}_{\rm HI} (1+ \beta\, \mu^2) - b_c f_V W(kR) \right]^2  P_{m}(k,z) \\
    &+ \bar{T}^2(z) \frac{f^2_V W^2(k R) }{\bar{n}_{\rm HI}}~,
\end{split}
\label{eq:hipk}
\end{equation}
where $W(kR)=3[\sin(kR) - kR \cos(kR)]/(kR)^3$ is the Fourier transform of the spherical top hat window function, $\mu$ is the cosine of the angle between $\kvec$ and the LoS vector $\n$, the term $(1+\beta \mu^2)$ arises due to the peculiar motion of the \HI~atoms and $\beta$ denotes the linear redshift distortion parameter \citep{bharadwaj2001,Bharadwaj_2004}. The first term which contains dark matter power spectrum $P_{m}(k)$ arises from the clustering of the hydrogen and the clustering of the centers of the ionized spheres. The second term which has $1/\bar{n}_{\rm HI}$ arises due to the discrete nature of the ionized regions.

Figure \ref{fig:eorpk} shows the behaviour of $P(k)$ as a function of $k$ at  three different frequencies $110 \,{\rm MHz}$, $120 \,{\rm MHz}$ and   $130 \,{\rm MHz}$ which corresponds to redshift $\sim 11.9, 10.8$ and $9.9$ respectively. At $\nu \le 110 \,{\rm MHz}$ the \HI~power spectrum traces the dark matter power spectrum as the hydrogen is largely neutral. The results shown here assume $b_{\rm HI}=2$ and the centers of the ionized bubbles to be clustered with a bias $b_c=1.5$ with respect to the underlying matter distribution. The presence of these clustered bubbles reduces the signal. At $130 \, {\rm MHz}$ \HII~bubbles occupy around $50 \%$ of the universe (Figure \ref{fig:nf}). The clustering of the bubbles or the dark matter is no longer important at this stage, and the \HI~power spectrum is largely governed by the discrete nature (Poisson distribution) of the ionized  bubbles. The power spectrum drastically falls towards the large $k$ which corresponds to the scales smaller than the typical bubble sizes. At lower redshifts (higher frequencies) the \HI~signal is expected to decreases rapidly because most of the \HI~would be ionized. The analytical model for the reionization breaks down beyond a stage when a large fraction of the universe is ionized. This break down occur beyond $130 \, {\rm MHz}$ for the model presented here. Note that the power spectrum shows oscillations for smaller length-scales $(k > \pi/R)$ which arises from the nature of the window function $W(kR)$ (eq. \ref{eq:hipk}). At these scales, the amplitude of $W^2(kR)$ falls as $(kR)^{-4}$ which is steeper than the matter power spectrum $P_{m}(k)$. Hence, at sufficiently large $k$ the EoR power spectrum $P(k)$ is dominated by the dark matter fluctuations and it approaches $P_{m}(k)$, with the approach being faster for large $R$. It is important to note that the oscillations in $P(k)$ are due to the fact that all the ionized bubbles are assumed to be spheres of the same size. In reality, the ionized regions will have a spread in the bubble shapes and sizes.

However this model has a limitation that it cannot be used when a large fraction of the volume is ionized as the ionized spheres start to overlap and the \HI~density contrast becomes negative in the overlapping regions. Calculating the fraction of the total volume where the \HI~density contrast is negative, is found to be $f_V^2/2$. Hence the model is valid for $z > 10$ where $f_V < 0.5$, and the $\delta_{\rm HI}$ is negative in less than $12.5\%$ of the total volume.

In the post-reionization era ($ z \le 6 $), the bulk of the \HI~resides in the high column density clouds which produce the damped Lyman-$\alpha$ absorption lines as observed in the quasar spectra \citep{lanzetta,lombardi,peroux}. The current observations indicate that the comoving density of \HI~expressed as a fraction of the present critical density is nearly constant at a value $\Omega_{\rm HI}(z) \sim 10^{-3}$ for $z \ge 1$ \citep{peroux,not,zafar}. The damped Lyman-$alpha$ clouds are believed to be associated with galaxies which represent highly non-linear overdensities. However, on the cosmological large-scales, it is reasonable to assume that these \HI~clouds trace the dark matter with a constant, linear bias $b_{\rm HI}=2$ \citep{Deb16}. 

Converting $\Omega_{\rm HI}$ to the mean neutral fraction $\xb=\Omega_{\rm HI}/\Omega_{\rm b}$ gives us $\xb=50 \Omega_{\rm HI}  h^2 (0.02/\Omega_{\rm b} h^2)$ or $\xb=2.45 \times 10^{-2}$. It is  also assumed that $T_{\rm s} \gg T_{\gamma}$, and hence one sees the \HI~in emission. Using these we have 

\begin{equation}
P(\kvec,z)= \bar{T}^2(z)\,b^2_{\rm HI}\, \bar{x}^2_{\rm HI}\left( 1+ \beta  \mu^2 \right)^2
\,P_m(k ,z)~.
\label{eq:postpk}
\end{equation}
The fact that the neutral hydrogen is in discrete clouds makes a contribution which is not included here. Another important effect not included here is that the fluctuations become non-linear at low $z$. Both these effects have been studied using simulations \citep{bharadwaj2004}.

\begin{figure}[ht!]
\centering
\includegraphics[width=0.5\textwidth, angle=0]{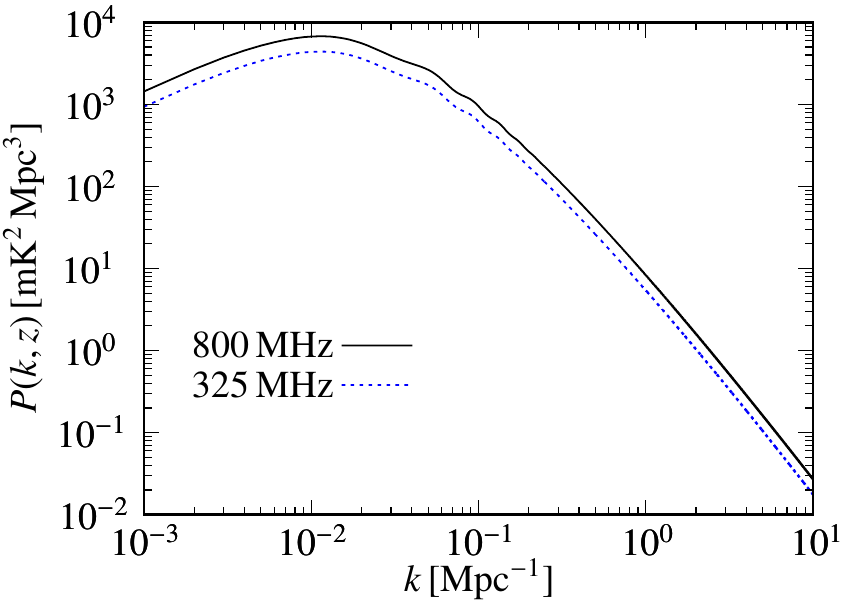}
\caption{Shows the expected post-reionization \HI~power spectrum $P(k, z)$ at two frequencies (as shown in the figure).}
\label{fig:peorpk}
\end{figure}

The predictions for post-reionization \HI~power spectrum $P(k)$ are shown as a function of $k$ for two frequencies $325 \,{\rm MHz}\, (z \approx 3.36) $ and $800 \,{\rm MHz}\, (z \approx 0.77)$ in Figure \ref{fig:peorpk}. The \HI~in this era is assumed to trace the dark matter field with a bias $b_{\rm HI}=2$. Therefore, the shape of the \HI~power spectrum as a function of $k$ is decided by the dark matter power spectrum $P_{m}(k)$ at the relevant Fourier modes. The values of $P(k)$ increase with frequency (decrease with redshift) is a reflection of the fact that $P_{m}(k)$ grows with time.


\subsection{Numerical Modeling}\label{subsec:numeric}
Even though analytical models provide some approximate pictures of CD-EoR 21-cm signal, it is not possible to incorporate various complex physics into them. It is thus necessary to use numerical simulations to obtain realistic maps of CD-EoR 21-cm signal and other quantities relevant to the observations. The numerical models typically stands on basic assumption that the hydrogen traces the underlying dark matter field and the collapsed halos host the ionizing sources.

\subsubsection{Radiative Transfer Technique}
\label{sec:RT}

\begin{figure*}
\centering
\includegraphics[width=0.95\textwidth, angle=0]{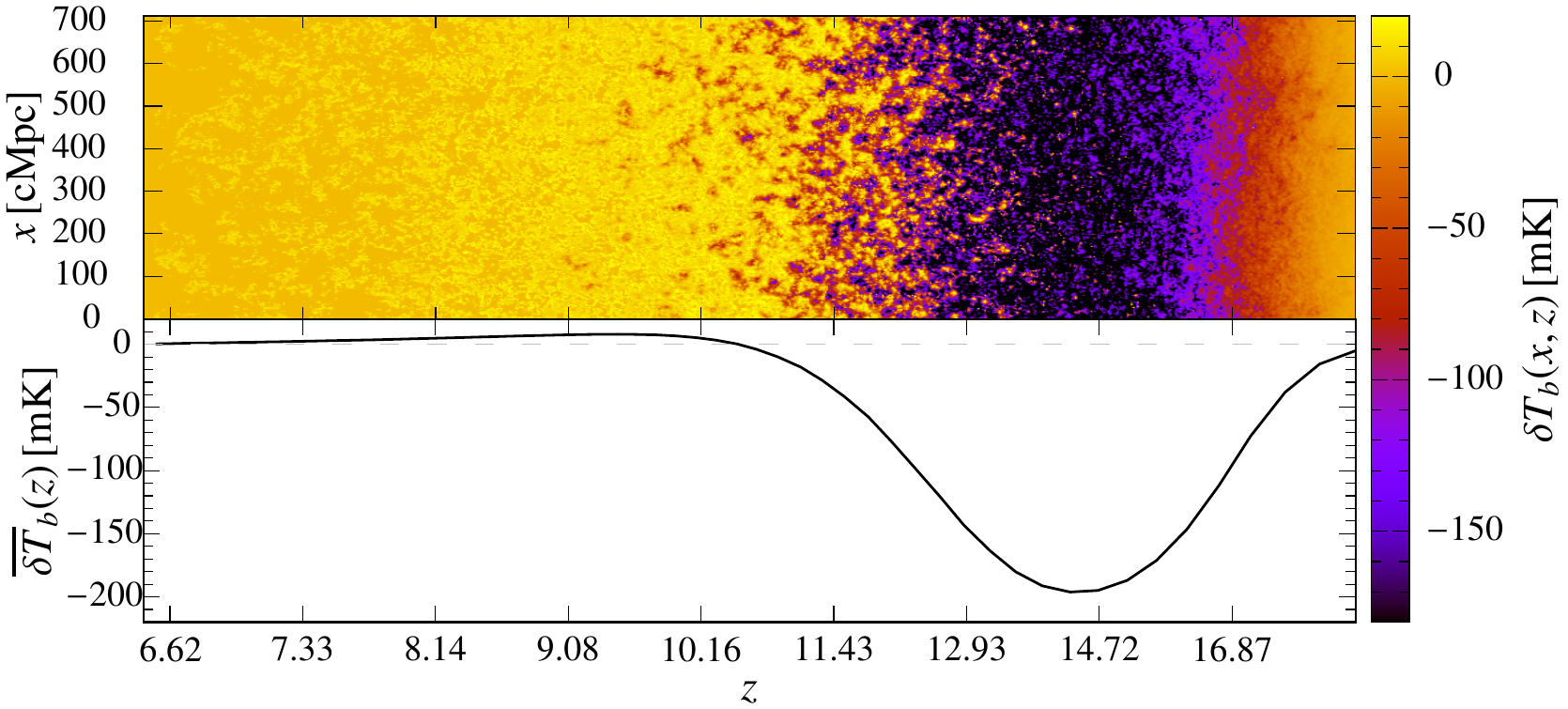}
\caption{The top panel shows a light-cone demonstrating the redshift evolution of the brightness temperature fluctuations of the \HI~21-cm signal during CD-EoR phase. The bottom panel shows the evolution of the corresponding mean brightness temperature of the 21-cm signal. The details of the dark matter only \textit{N}-body simulation as well as the source model used in {\sc GRIZZLY} to generate this light-cone can be found in \citet{ghara15b}.}
\label{fig:GRIZZLY_LC}
\end{figure*}

We have already seen that the intensity of the \HI~21-cm radiation crucially depends on the ionization fraction, gas temperature and the Lyman-$\alpha$ coupling (see eqs. \ref{eq:bt} and \ref{eq:Ts}). Thus, besides knowing how the radiating sources such as galaxies, high-mass X-ray binaries, Quasars are distributed in the simulation volume and their spectral energy distribution (SED), it is also important to know how the emitted UV, X-ray and Lyman-$\alpha$ photons propagate into the clumpy baryonic field. One can simplify the picture of star formation by assuming that the stellar content of a galaxy is proportional to the mass of the host dark matter halo and SEDs of the galaxies per unit stellar mass are the same as one can obtain from a stellar population synthesis code like {\sc pegase2} \citep{Fioc97}. Note that, in reality, the star formation scenario can be very complex. 

In principle, one needs to solve a seven-dimensional cosmological radiative transfer equation (see e.g., \cite{2016JApA...37...29C}) at every grid point of the simulation box to include all relevant physical processes. This is computationally very challenging and thus, often approximations are used for the radiation transfer. For example, a 3D radiative transfer code C$^2${\sc ray} \citep[][]{mellema06} works by tracing rays from all UV emitting sources and iteratively solving for the time evolution of $x_{\rm HII}$. Even with such approximation, C$^2${\sc ray} results match very well with fully numerical radiative transfer Codes \citep[see e.g.,][]{Iliev2006}. Alternatively, a one-dimensional radiative transfer scheme is used in codes such as {\sc GRIZZLY} \citep{ghara15a, ghara15b} and {\sc BEARS} \citep{Thom08}.  This method approximates the transfer of UV, X-rays, and Lyman-$\alpha$ photons by assuming that the effect from individual sources is isotropic. Although {\sc GRIZZLY} employs a range of approximations to correct for overlaps between individual ionized regions around the sources, its results are still in good agreement with those of the 3D radiative transfer code C$^2${\sc ray} \citep[see][for details]{ghara18}, while being at least $10^5$ times faster. In addition, as {\sc GRIZZLY} also includes X-ray heating and Ly-$\alpha$ coupling, it can probe the cosmic dawn where the spin temperature fluctuations are expected to dominate the fluctuations in the 21-cm signal. We present one simulated {\sc GRIZZLY} light-cone of the brightness temperature in Figure \ref{fig:GRIZZLY_LC}. Clearly, at the beginning of the Cosmic Dawn era (around $z\sim20$ in this case), Lyman-$\alpha$ coupling is the most crucial effect that determined the fluctuations in the 21-cm signal. This was followed by an era where X-ray heating becomes important. One can see that over time individual emission regions around the X-ray sources overlap together. The gas in the IGM became significantly heated around $z\sim 12$ after which spin temperature fluctuations became negligible compared to the ionization fluctuations. 

\begin{figure*}
	\centering
	\includegraphics[width=1\textwidth, angle=0]{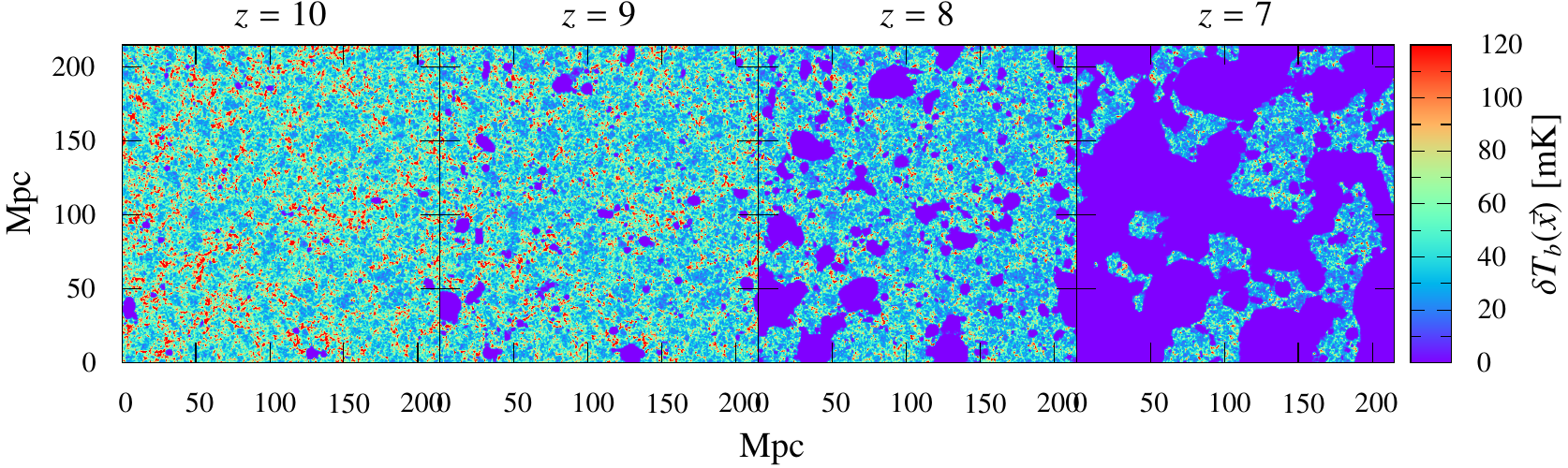}
	\caption{Shows one realization of the two dimensional sections through the simulated \HI\, brightness temperature maps at $z = 10$, $9$, $8$ and $7$ within simulation volume of $[215\, {\rm Mpc}]^3$. See \citet{Mondal2017} for details.}
	\label{fig:HI_map}
\end{figure*}


As mentioned before, radiative transfer methods are based on post-processing of the outputs of \textit{N}-body simulations (such as the dark matter halo catalog, the density and velocity fields, etc), these miss the hydro-dynamical effects on the small scales. On the other hand, more sophisticated fully coupled radiative-hydrodynamic simulations of galaxy formation such as Renaissance \citep{2016ApJ...833...84X}, CoDa I \citep{2018ApJ...856L..22A}, CoDa II  \citep{codaII} by CPU-GPU code EMMA \citep{2015MNRAS.454.1012A} can include several small scale physics and thus are useful to study the stochastic star formation, metal enrichment, radiative feedback, recombination in the IGM, etc. The impact of such relevant processes as found from these studies are often approximately used in other types of simulations such as semi-numerical and numerical without hydrodynamics. While these simulations are very important to consider small-scale physical processes, these are often limited by small box size and thus challenging to compare with the 21-cm observations. In addition, resolving very fine scale physical processes such as a supernova, turbulence, etc. are still beyond the capacity of these methods and often depends on simplified sub-grid prescriptions.


\subsubsection{Semi-numerical simulations}\label{subsec:sem-num}

Detailed and fast models of reionization are a crucial ingredient for interpreting the observed data and constraining the physics of the CD-EoR. The aforementioned radiative transfer simulations are required for accurate modelling, however, they are not well-suited for exploring a vast range of parameter space \citep[see e.g.][]{Mondal2020b}. Therefore, fast methods are crucial to simulate a high dynamic range large-volume boxes of the \HI~21-cm signal. This can be achieved by using semi-numerical techniques, which typically involves three main steps -- (a.) simulating the dark matter distribution at the desired redshifts, (b.) identifying the collapsed dark matter halos and (c.) generating the reionization map using an excursion set formalism. The first two steps mostly overlap with those in the Radiative transfer techniques. It's the excursion set \citep{Furlanetto2004} or similar formalism which generates ionization maps based on some approximate methodology.

For example, a semi-numerical code {\sc{ReionYuga}}\footnote{\url{https://github.com/rajeshmondal18/ReionYuga}} \citep{Mondal2015, Mondal2016, Mondal2017} generates ionization fields following methodologies described in \citet{Choudhury2009} and \citet{Majumdar2014}. The code uses the matter density field simulated using a fast, efficient and accurate parallelized particle-mesh $N$-body code\footnote{\url{https://github.com/rajeshmondal18/N-body}} \citep{Mondal2015}, and the corresponding halo distribution obtained using a Friends-of-Friends halo finder\footnote{\url{https://github.com/rajeshmondal18/FoF-Halo-finder}} \citep{Mondal2015}. The methodology of {\sc{ReionYuga}} assumes that the amount of ionizing photons produced by a source is directly proportional to its host halo mass above a certain mass cut-off $M_{\rm min}$ which is a parameter in simulations. The proportionality factor here is quantified through a dimensionless parameter $N_{\rm ion}$ which is related to several degenerate astrophysical factors \citep[see e.g.][]{Shaw2020}. The mean free path of the ionizing photons in the IGM $R_{\rm mfp}$ is the third parameter of {\sc ReionYuga}. The methodology determines whether a grid point to be ionized or not by smoothing the hydrogen density field and the photon density field using spheres of different radii starting from the grid size to $R_{\rm mfp}$. A grid point is considered to be ionized if for any smoothing radius (within $R_{\rm mfp}$) the photon density exceeds the hydrogen density at that grid point. Grid points which do not satisfy this condition are assigned with an ionized fraction. Figure~\ref{fig:HI_map} shows one realization of the two dimensional sections through the simulated \HI~brightness temperature maps at $z = 10$, $9$, $8$ and $7$ for a $[215\, {\rm Mpc}]^3$ simulation box. The resultant ionization maps are similar to those simulated using the costlier radiative transfer methods. 
 
There are several semi-numerical codes available publicly such as {\sc SCRIPT}\footnote{\url{https://bitbucket.org/rctirthankar/script}} \citep{script}, {\sc 21cmFAST}\footnote{\url{https://github.com/21cmfast/21cmFAST}} \citep{21cmfast} and {\sc SimFast21}\footnote{\url{https://github.com/mariogrs/Simfast21}} \citep{simfast21} etc. However these codes have their own assumptions and physically motivated parameters. One should appropriately choose these codes based on the individual requirement.


\section{Statistical measures of the 21-cm signal}\label{sec:stats}
The CD-EoR 21-cm signal (see eqs. \ref{eq:bt} and \ref{eq:dTb}) is a random field of brightness temperature fluctuations. We need  statistical estimators to quantify the signal. One can quantify it using the one-point statistics such as variance \citep{patil2014}, skewness and kurtosis \citep{harker2009} of $\delta T_b$. Despite being easier to compute and interpret, one-point statistics have limited information. Therefore many-point statistics, such as $N$-point correlation functions, are needed to quantify the signal completely. We shall discuss about a few many-point statistics in Fourier space such as power spectrum, bispectrum etc. in the following sections. One can also utilize other techniques, such as the matched filtering, to enhance the detectability of the signal by extracting its important underlying features.

\subsection{Matched filtering technique}
\label{sec:MF}

The detectability of the CD-EoR 21-cm signal in terms of different statistical quantities is low as the targeted signal is weak. Thus, one of the major focuses of the theoretical study of the signal has been to design optimum estimators that enhance the detectability of the signal. The application of the matched filtering technique which applies pre-defined filters to the measurements to enhance the detectability of the observation is one such complementary method.

\begin{figure}[ht!]
	\centering
	\includegraphics[scale=0.92]{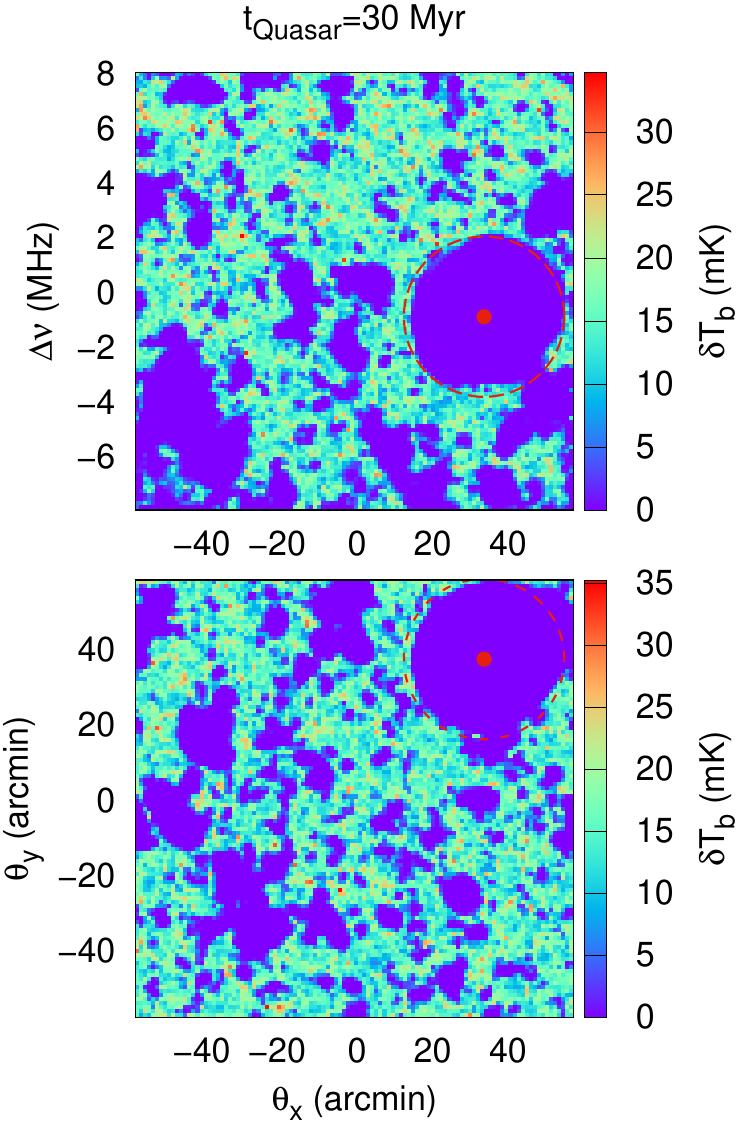} 
	\caption{Shows simulated $\delta T_b$ maps at $z \approx 7$ using {\sc grizzly}, the red points indicate the position of a Quasar in the field-of-view,  the dashed circles represent the best-fit size of matched-filter estimated from a Bayesian analysis as used in \citet{2020MNRAS.496..739G}.}
	\label{fig:MFil}
\end{figure}

In the case of the EoR 21-cm signal, \citet{Kanan_2007} introduced this method where spherical top-hat filters in the image-space were applied to the measured visibilities for enhancing the detectability of ionized regions. Follow up studies such as \citet{2008MNRAS.391.1900D, 2009MNRAS.399L.132D, datta2012a, 2011MNRAS.413.1409M, 2012MNRAS.426.3178M} considered various scenarios of IGM ionization states and estimated the detectability of ionized bubbles with radio interferometers like GMRT, MWA, LOFAR, SKA-Low. For example, \citet{datta2012a} showed that $\sim 1200$ hours of observation with LOFAR should be enough to detect ionized regions with a size $\sim 25\,{\rm cMpc}$ at redshift $7.57$ with mean ionization fraction $\xb=0.5$. As the largest ionized regions in the field-of-view (FoV) is expected to be formed around very bright sources such as high-$z$ Quasars, this technique can also give information about the bright sources \citep[see e.g.,][]{datta2012a, 2012MNRAS.426.3178M}.

Recently, \citet{2020MNRAS.496..739G} developed a rigorous Bayesian framework based on the same method to constrain the parameters that characterize the ionized regions. This study shows $20$ hours integration with SKA-Low will be enough to constrain the location and size of the ionized regions around typical quasars at $z=7$ (see Figure \ref{fig:MFil}), while it also constrains the difference in the neutral fractions inside and outside the \HII~bubble. This framework is useful in identifying large ionized regions in the observed field and hence will be interesting for following up observations with deeper integration times. 

In contrast to the distribution of ionized regions in the IGM during the EoR, the CD \HI~21-cm signal mostly depends on how emission/absorption regions are distributed in the IGM. In this case, one can also apply the matched filtering technique to enhance the detectability of individual emission/absorption regions in the IGM. In a slightly different approach, one can also simply combine the measured visibility from different baselines and frequency channels to enhance the detectability of the signal \citep[see e.g.,][]{ghara16a}. We refer the reader to \citet{2016JApA...37...27D} for more details on these methods. 


\subsection{3D Power Spectrum}\label{subsec:2.3.1}

\begin{figure}[ht!]
	\centering
	\includegraphics[width=0.5\textwidth, angle=0]{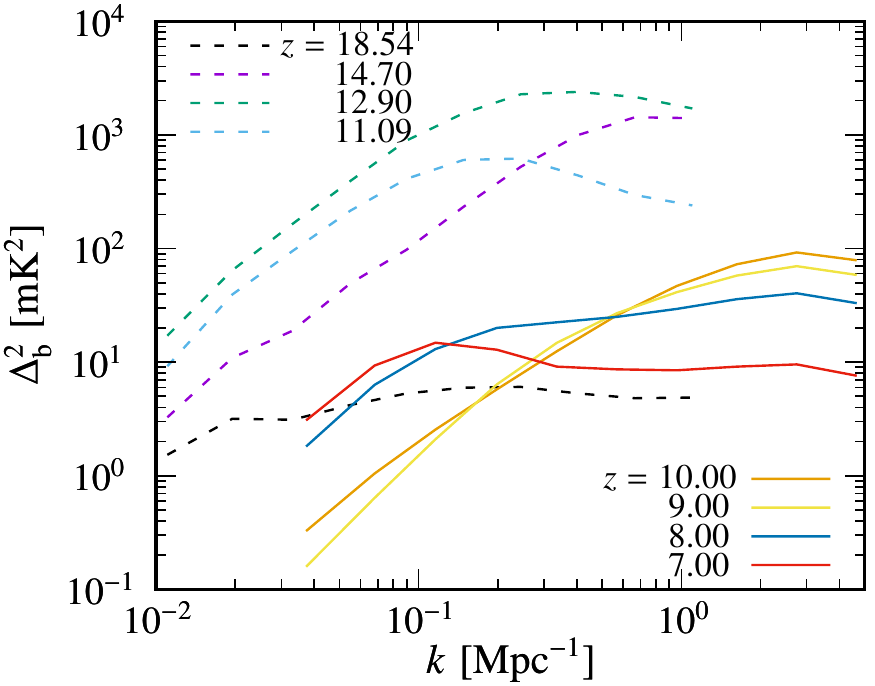}
	\caption{Shows the dimensionless form of CD (dashed lines) and EoR (solid lines) 21-cm 3D power spectrum $\Delta_{\rm b}^2(k) = k^3 P(k)/(2\pi^2)$. The CD power spectra are obtained from the radiative transfer simulations \textsc{GRIZZLY}, whereas the EoR power spectra are computed using the semi-numerical simulations \textsc{ReionYuga}. See \citet{ghara15b} and \citet{Mondal2017} for details.}
	\label{fig:pk}
\end{figure}

The spatial fluctuations in the \HI~differential brightness temperature $\delta T_b(\xvec)$ within a 3D comoving cube of volume $V$ can be decomposed into different Fourier modes in $\kvec$-space as
\begin{equation}
\Delta \Tilde{T}_b(\kvec) = V^{-1}\sum^{}_{\xvec} \exp[-i \kvec \cdot \xvec] \,\,\delta T_b(\xvec)~, 
\label{eq:eta}
\end{equation}
where $\Delta \Tilde{T}_b(\kvec)$ is the Fourier conjugate of $\delta T_b(\xvec)$ and the summation is over all real space points within the volume. Note that the wave vector $\kvec$ can be both positive and negative, however we have $\Delta \Tilde{T}^{*}_b(\kvec) = \Delta \Tilde{T}_b(-\kvec) $ as $\delta T_b(\xvec)$ is a real field. Now the \HI~21-cm 3D power spectrum $P(\kvec)$ can be defined as follows
\begin{equation}
    \langle \Delta \Tilde{T}_b(\kvec_1) \Delta \Tilde{T}_b(\kvec_2)\rangle= V\, \delta^{\rm K}(\kvec_1+\kvec_2) \,P(\kvec_1)~,
\label{eq:PHI}    
\end{equation}   
where $\delta^{\rm K}$ is the 3D Kronecker's delta function. The power spectrum is the Fourier pair to the two-point correlation function, which measures excess probability as a function of two points in real space. We now discuss a few important properties of the $P(\kvec)$. The 3D Kronecker's delta function $\delta^{\rm K}$ appearing in above eq. (\ref{eq:PHI}) is a consequence of the fact that the \HI~fluctuations are assumed to be statistically homogeneous or ergodic in nature along all three spatial directions. It signifies that the \HI~signal in the different Fourier modes are uncorrelated. Assuming statistical isotropy in the signal removes the direction dependence of the 3D power spectrum, and we can express power spectrum as a function of only $k$ instead of $\kvec$, \textit{i.e.} $P(\kvec)=P(k)$. In reality, the CD-EoR 21-cm signal is affected by several LoS effects such as the redshift-space distortion, light-cone effect, Alcock-Paczynski effect. However, the statistical isotropy is always valid for the monopole component of the 3D power spectrum $P(k)$. Figure \ref{fig:pk} shows dimensionless 3D power spectrum $\Delta_{\rm b}^2(k) = k^3 P(k)/(2\pi^2)$ of the EoR 21-cm signal (solid lines) obtained from an ensemble of simulated signals at four redshifts as shown in Figure \ref{fig:HI_map}. In Figure \ref{fig:pk}, we also show the CD 21-cm 3D power spectrum (dashed lines) which have been simulated using the 1D radiative transfer simulation {\sc GRIZZLY} as mentioned in section \ref{sec:RT}.

One of the primary goals of the ongoing radio-interferometric experiments is to measure the 3D power spectrum of the EoR 21-cm signal. However these interferometers are limited in terms of sensitivity. The SKA-Low in its first phase is expected to provide EoR 21-cm power spectrum at a considerably high SNR level within $100$ hours of observations \citep{Shaw2019}. Owing to its larger frequency bandwidth, SKA-Low will be able to provide measurements of CD 21-cm power spectrum at a reasonably high SNR. 


\subsection{Multifrequency Angular Power Spectrum}
\begin{figure*}
	\centering
	\includegraphics[width=.99\textwidth, angle=0]{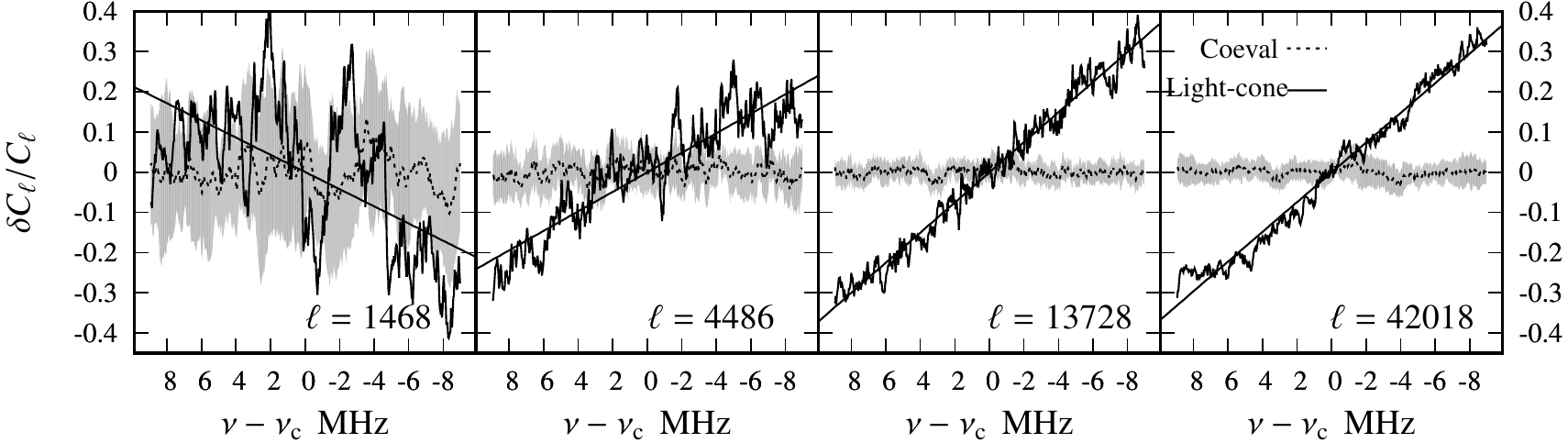}
	\caption{The information loss in the assumption of statistical homogeneity and periodicity along the frequency direction. Dotted curves are for the power spectrum which assumes the signal is ergodic and periodic along the frequency direction and the soiled curves are for MAPS which does not require these assumptions. See \citet{Mondal2018} for details.}
	\label{fig:delta_cl}
\end{figure*}

The CD-EoR 21-cm signal is observed across a frequency bandwidth for a range of cosmic times, as each wavelength corresponds to a different look-back time. The observed data sets are thus three-dimensional (3D) with the two directions on the sky-plane and the frequency (wavelength) constituting the third dimension. The light-cone effect imprints the cosmological evolution of the 21-cm signal along the LoS direction (or frequency axis) \citep{datta2012}. The effect is particularly pronounced when the mean 21-cm signal changes rapidly as the universe evolves. \citet{Mondal2018} have developed a method to properly incorporate the light-cone effect in simulations of the 21-cm signal that also includes the effects of peculiar velocities. They showed that the 3D power spectrum $P(k)$ fails to quantify the entire two-point statistical information as it inherently assumes the signal to be ergodic and periodic in all three directions (Figure\,\ref{fig:delta_cl}), whereas the light-cone effect breaks these conditions along the LoS direction \citep{Mondal2018}. Therefore, the issue is how to analyse the statistics of the 21-cm signal in the presence of the light-cone effect. \citep{datta2012} and \citet{Mondal2018} have proposed an unique statistical estimator of the signal, the multifrequency angular power spectrum (MAPS), which quantifies the entire second-order statistics of the 21-cm signal without requiring the signal to be ergodic and periodic along the LoS direction. 

The Multifrequency Angular Power spectrum (MAPS) is defined as the following
\begin{equation}
C_{\ell} (\nu_1, \nu_2) = \Omega^{-1} \, \langle \Delta \tilde{T}_{b2} (\u, \nu_1) \, \Delta \tilde{T}_{b2} (-\u, \nu_2) \rangle ~,
\end{equation}
for a given multipole $\ell = 2 \pi |\u|$ and at a pair of frequencies $(\nu_1, \nu_2)$. In the above equation, $\Omega$ is the solid angle subtended by the observation volume to the observer, $\Delta \tilde{T}_{b2}(\u, \nu)$ is the Fourier conjugate of the \HI~21-cm brightness temperature fluctuations $\delta T_b(\vec{\theta}, \nu)\equiv \delta T_b({\xvec})$ with respect to the two-dimensional angle $\vec{\theta}$ defined on the sky-plane, and $\u$ is the Fourier conjugate of $\vec{\theta}$. For the case where the signal is assumed to be both ergodic and periodic, the MAPS bears a direct relation to the three-dimensional power spectrum $P(\kvec)$ (see sec.~\ref{subsec:2.3.1}), \textit{i.e.} 
\begin{equation}
C_{\ell}^{\rm EP} (\Delta \nu) = (r^2_c r^{'}_c B)^{-1} \, \sum_{k_{\parallel}} \, {\rm e}^{i k_{\parallel} r^{'}_c \Delta \nu} \, 
P(\kvec_{\perp}, k_{\parallel})\,,
\label{eq:cl_ep}
\end{equation}
where $\Delta \nu = |\nu_1 - \nu_2|$, $r_c$ and $r^{'}_c$ denote the distance between the observer and the center of the observation volume and its derivative w.r.t. $\nu$ and $B$ gives the observing bandwidth. In eq. (\ref{eq:cl_ep}), $\kvec_{\perp}$ and  $k_{\parallel}$ respectively denote the perpendicular and parallel components of the 3D wavevector $\kvec$ w.r.t. the LoS direction of the observer.  

\citet{Mondal2020} have applied MAPS to quantify the statistics of light-cone EoR 21-cm signals and study the prospects for measuring it using the upcoming SKA observations. In a previous study \citep{Mondal2019}, the authors have also found that it is possible to recover the cosmic history across the entire observational bandwidth using MAPS statistics. Furthermore, by construction the MAPS is more natural to use for radio-interferometric observations than the 3D power spectrum $P(k)$ which provides biased estimates of the signal \citep{Mondal2020}.


\subsection{Higher order statistics}
\label{sec:nonGaussian}

\begin{figure}[ht!]
    \centering
    \includegraphics[width=0.4\textwidth]{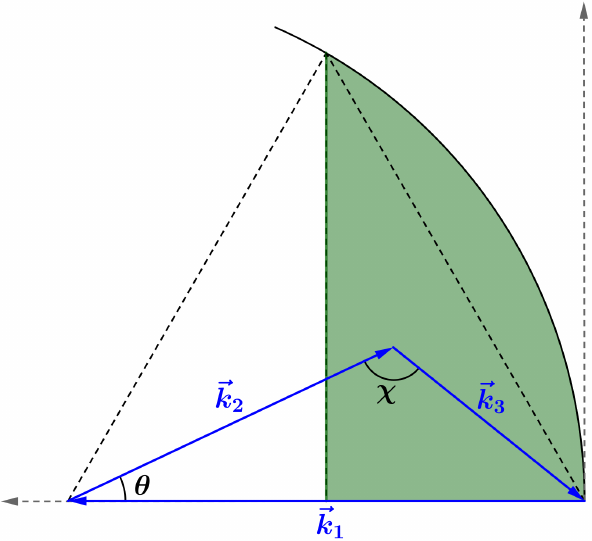}
    \caption{Shows the definition of triangles in $\kvec$-space. The shaded region shows the positions of all the unique point of intersections of $\kvec_1$ and $\kvec_2$ vectors, while satisfying eqs. (\ref{eq:k_relation})--(\ref{eq:costheta}).}
    \label{figs:uni_trig}
\end{figure}

The two-point Fourier space statistics (\textit{i.e.} power spectrum) can provide a complete statistical description of a pure Gaussian random field. However, the fluctuations in the CD-EoR 21-cm signal $\delta T_b(\xvec)$ are highly non-Gaussian \citep{bharadwaj05a,mellema06}. This non-Gaussianity arises due to the non-random distribution of the first luminous sources in the IGM and the interaction of radiation from these sources with IGM gas through various complex astrophysical processes such as the heating and ionization during CD-EoR. Furthermore, this intrinsic non-Gaussianity evolves (either increases or decreases) with time as the heating and the ionization fronts grow and percolate in the IGM. It is necessary to consider the non-Gaussian statistics of the signal while interpreting it thoroughly via statistical inference methods. As discussed in section \ref{subsec:2.3.1}, the power spectrum $P(k)$ is a resultant of signal correlation at a single Fourier mode, and therefore it is not sensitive to the non-Gaussianity present in the CD-EoR 21-cm signal. To capture this inherent non-Gaussianity present at various length scales, one needs the higher-order statistics such as the bispectrum (three-point), trispectrum (four-point), etc. The bispectrum, the Fourier transform of the three-point correlation function, is the lowest order statistic that can directly probe the non-Gaussianity present in the signal. We define bispectrum $B(\kvec_1, \kvec_2, \kvec_3)$ as

\begin{equation}
\begin{split}
    \langle \Delta \Tilde{T}_b(\kvec_1) \Delta \Tilde{T}_b(\kvec_2)
 \Delta \Tilde{T}_b(\kvec_3)\rangle = V\, \delta^{\rm K}&(\kvec_1+ \kvec_2+ \kvec_3)\\
 &\times B(\kvec_1, \kvec_2, \kvec_3) \,,
\end{split}
\label{eq:bispec}
\end{equation}
where $\Delta \Tilde{T}_b(\kvec)$ is the Fourier transform of $\delta T_b(\xvec)$, and $\delta^{{\rm K}}(\kvec_1+ \kvec_2+ \kvec_3)$ is the Kronecker's delta function. The occurrence of Kronecker's delta function is due to statistical homogeneity of the signal. This indicates that bispectrum is defined only when the three vectors ($\kvec_1, \kvec_2, \kvec_3$) form a closed triangle in $\kvec$-space, \textit{i.e.} $\kvec_1 +\kvec_2 +\kvec_3 = 0$, shown in Figure \ref{figs:uni_trig}. The bispectrum, hence provides us with the correlations among the signal at different Fourier modes and thus is capable to capture the non-Gaussianity.

Recently, \citet{trott19} for the first time tried to put an upper limit on the 21-cm bispectrum using the observational data from the MWA Phase II array. The detection of the CD-EoR 21-cm bispectrum will require more sensitive observations than needed for power spectrum detection. The next-generation radio interferometers such as HERA and upcoming SKA-Low are expected to have enough sensitivity for bispectrum measurements. It is thus more relevant now to predict the characteristics of the CD-EoR 21-cm bispectrum in the view of the construction of SKA-Low.

\begin{figure*}
    \centering
    \includegraphics[width=0.9\textwidth]{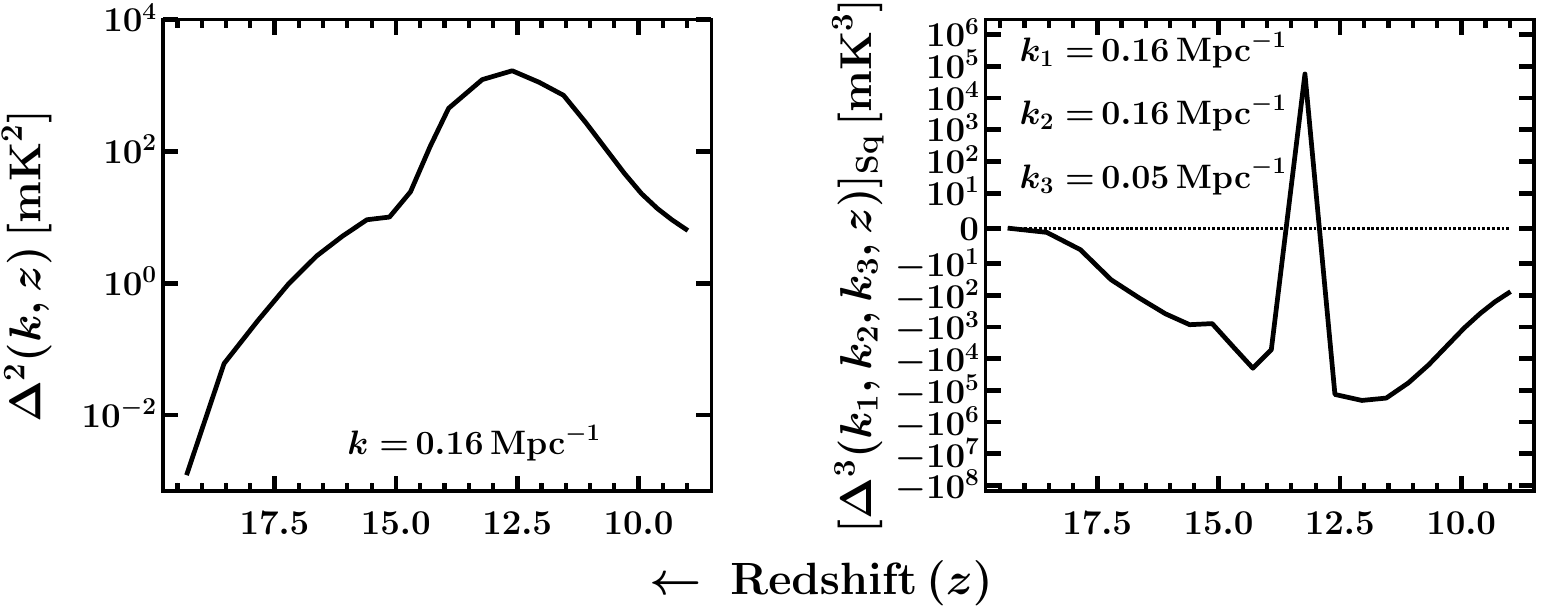}
    \caption{{\it Left panel}: Redshift evolution of the 21-cm power spectrum during CD at $k = 0.16\, {\rm Mpc}^{-1}$. {\it Right panel}: Redshift evolution of the 21-cm bispectrum for squeezed-limit $k$-triangle during CD at $k_1 = 0.16\, {\rm Mpc}^{-1}$($k_2 = 0.16\, {\rm Mpc}^{-1}$, and $k_3 = 0.05\, {\rm Mpc}^{-1}$).}
    \label{fig:PSBS_z}
\end{figure*}

Several studies on the CD-EoR 21-cm bispectrum have been performed so far. \citet{bharadwaj05} and \citet{ali05} made the first theoretical attempts to characterize the EoR 21-cm bispectrum using their analytical models. One of the most crucial features of the bispectrum predicted in their studies is its sign that can attain both positive and negative values. Later on, a number of independent studies, such as \citet{watkinson17,majumdar18,majumdar20,hutter19,watkinson19,kamran21,kamran22}, confirmed the sign of the CD-EoR 21-cm bispectrum using the detailed simulations of the 21-cm signal from the CD-EoR and for a variety of $k$-triangles (equilateral, isosceles and scalene). The sign of the bispectrum is an important feature of this statistic. Figure \ref{fig:PSBS_z} clearly demonstrates that the redshift evolution of the bispectrum during CD, for instance, has way more features in its magnitude and sign than the power spectrum evolution. The sign and magnitude of the bispectrum uniquely identify which one among the dominant physical processes such as the Ly$\alpha$ coupling, X-ray heating, and photo-ionization is driving the non-Gaussianity at a particular cosmic time. For further insights, we refer the interested reader to \citet{kamran22}. Furthermore, when analyzed, the sign change of the bispectrum's evolution with triangle configurations and the redshift can be used as a confirmatory test of the CD-EoR 21-cm mean signal detection.

A comprehensive view of the non-Gaussianity requires the study of the bispectra for all possible triangles in the Fourier space. \citet{bharadwaj20} for the first time, proposed a new parameterization for the bispectrum of triangles of all possible unique shapes in the Fourier space, shown by the shaded region in Figure \ref{figs:uni_trig}, while the following conditions are satisfied.

\begin{eqnarray}
&&{k}_1 \geq {k}_2 \geq {k}_3
\label{eq:k_relation}\\
&&0.5 \leq n \leq 1.0
\label{eq:n}\\
&& n \cos\theta \geq 0.5.
\label{eq:costheta}
\end{eqnarray}
Where, $n = k_2/k_1$ and $\cos{\theta} = -{\vec{k}}_1\cdot {\vec{k}}_2/{k_1 k_2}$.

\citet{majumdar20} and \citet{kamran21} followed the same prescription and presented a comprehensive and correct interpretation of the CD-EoR 21-cm signal by studying the bispectrum of all possible unique triangles using the simulated signal. These studies conclude that the squeezed limit bispectrum typically attains the largest magnitude among all possible unique triangles in the Fourier space. The squeezed-limit bispectrum is, thus, expected to have the largest detection probability.


\subsection{Intrinsic errors in statistical estimators}
The statistical errors which are inherent to the measurements of the CD-EoR 21-cm signal have two components -- (1) {\it cosmic variance}, which is inherent to the signal itself and (2) {\it system noise}, which is observationally inevitable. The cosmic variance error arises from the finite volume of the universe accessible to a measurement whereas the system noise is observed by the receivers even looking at a hypothetically blank sky. There have been several works to predict the errors on the EoR 21-cm power spectrum $P(k)$ measurements \citep[see e.g.][]{zaroubi2012,parsons12,jensen13,pober2014, Ewall-Wice_2016}. Most of these earlier works have commonly assumed the signal to be a Gaussian random field, for simplicity. 
However, the EoR 21-cm signal is a highly non-Gaussian field as already discussed in section~\ref{sec:nonGaussian}. This non-Gaussianity is expected to affect the total error in the measurement of any statistical estimator of the CD-EoR 21-cm signal. The system noise, on the other hand, is considered to be a Gaussian random field whose impact in the error covariance would be competing with the non-Gaussianity of the signal.

Considering the 3D power spectrum $P(\vec{k})$ (eq. \ref{eq:PHI}), the corresponding statistical error covariance between the measurements at $\vec{k}_i$ and $\vec{k}_j$ can be written as \citep[see e.g.,][]{Mondal2016,Shaw2019}
\begin{equation}
    \mathbf{C}_{ij} = [P^2_{\rm b}(\vec{k}_i) + P^2_{\rm N}(\vec{k}_i)]\, \delta^{\rm K}_{ij}+ \frac{T(\vec{k}_i, -\vec{k}_i,\vec{k}_j,-\vec{k}_j)}{V}\,.
    \label{eq:error_cv}
\end{equation}
Here the $P_{\rm b}$ and $P_{\rm N}$ respectively denotes the 21-cm PS and the system noise power spectrum. Note that $\mathbf{C}_{ij}$ has contribution from the four-point statistics, trispectrum $T(\vec{k}_i, -\vec{k}_i,\vec{k}_j,-\vec{k}_j)$, which would have been zero if the signal were a Gaussian random field.

\begin{figure}[ht!]
\centering
\includegraphics[width=0.5\textwidth, angle=0]{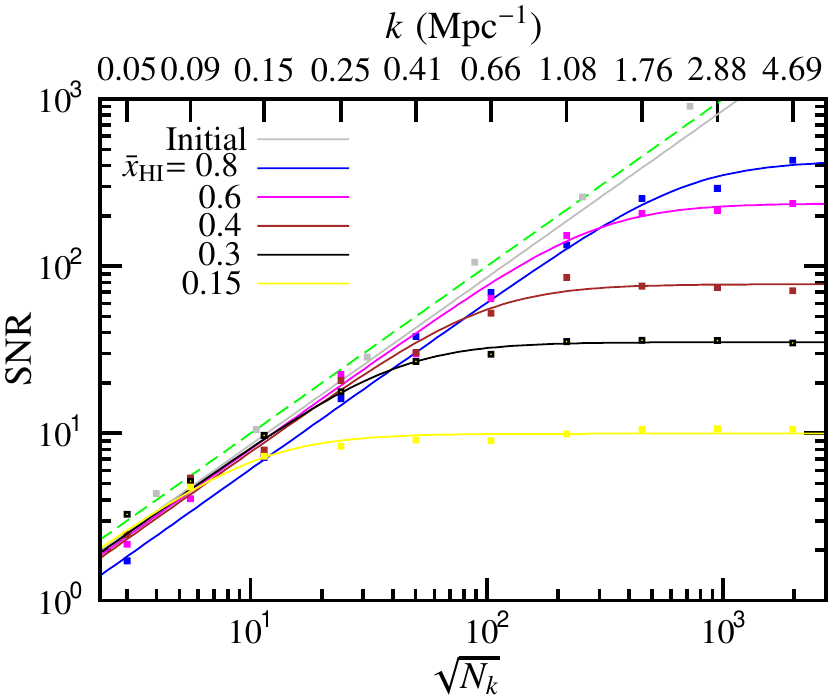}
\caption{Shows the SNR as a function of $\sqrt{N_k}$. The $45^{\circ}$ dashed line shows the SNR expected for a Gaussian random field, and `Initial' refers to the input linear density fluctuations used for the dark matter {\it N}-body simulations. The other data points are the SNR estimated for the different stages of reionization as mentioned by $\xb$ in the legend. Credit: \citet{Mondal2015}.}
\label{fig:snr_pk_cv}
\end{figure}

The trispectrum contribution in eq. (\ref{eq:error_cv}) is majorly a result of the formation of the ionized bubbles in the IGM. The non-zero trispectrum indicates that the ionized bubble formation is not a random process, therefore the different $k$ modes have correlated information. If the EoR 21-cm signals were a Gaussian random field the different $k$ mode would have been uncorrelated, and we expect the signal-to-noise ratio (SNR) for $P(k)$ to scale as the square-root of the number of Fourier modes $N_k$ that are being averaged over. However study of \citet{Mondal2015} show that the SNR saturates to a limiting value, even if one increases the number of Fourier modes (see Figure \ref{fig:snr_pk_cv}). In a follow up work, \citet{Mondal2016} presented a detailed and generic theoretical framework to interpret and quantify the effect of non-Gaussianity on the error estimates for $P(k)$ through the full error (cosmic) covariance matrix. Notably, \citet{Mondal2017} studied how the effect of non-Gaussianity on the reionization power spectrum cosmic variance evolves as the reionization progresses. These work collectively conclude that the effect of non-Gaussianity is dominant on the large $k$ modes (small length-scales) and also towards the later stages of reionization ($\xb \leq 0.5$) as seen in Figure \ref{fig:snr_pk_cv}. It is important to note that all these three works do not included system noise contributions in their error analysis.

\begin{figure}[ht!]
  	\centering
  	\mbox{\includegraphics[width=0.45\textwidth,angle=0]{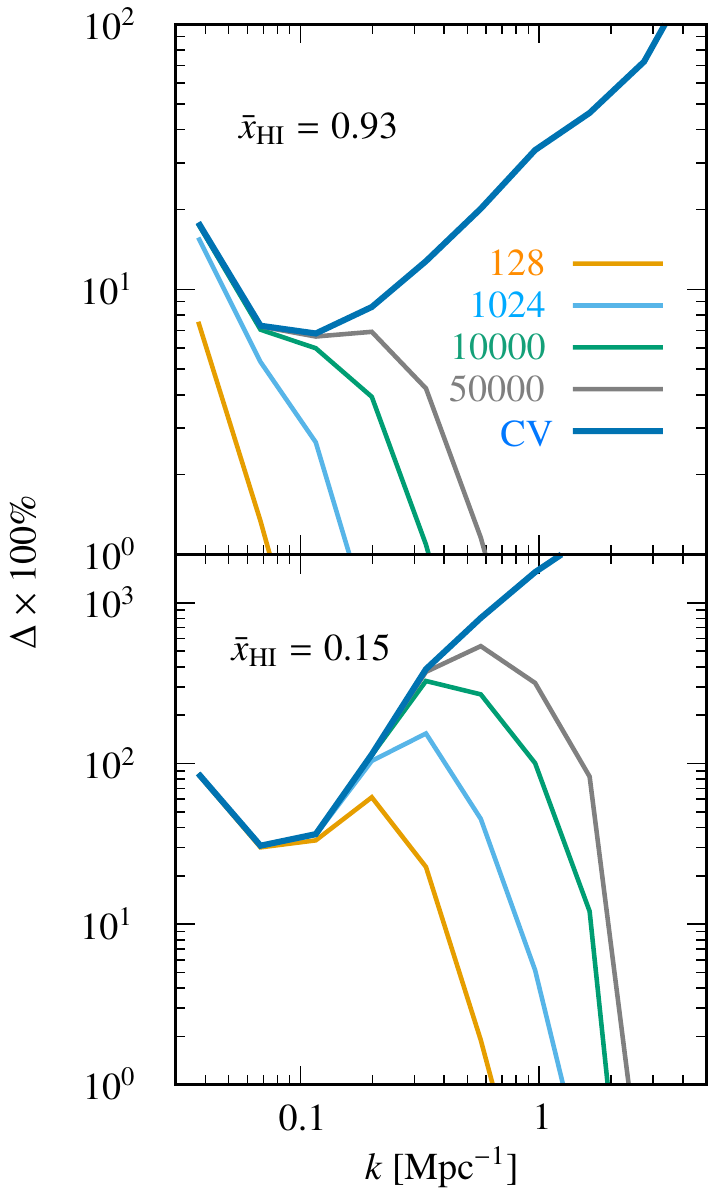}}
  	\caption{Shows the $\%$ deviation of non-Gaussian error variance in observed EoR 21-cm power spectrum with respect to the corresponding Gaussian predictions. Credit: \citep{Shaw2019}}
  	\label{fig:frac_1}
  \end{figure}

The large system noise contribution to the signal can wash out these non-Gaussian features in the observed CD-EoR 21-cm signal. The system noise power spectrum $P_{\rm N}$ \citep[see eq. (1) of][]{Shaw2019} scales inversely proportional to the observation time $t_{\rm obs}$ as the Gaussian system noise will reduce down if averaged over larger number of time stamps. $P_{\rm N}$ scales inversely also with the distribution of baselines $\vec{U}\propto \kvec_{\perp}$, \textit{i.e.} the $\kvec$ grids at which more baselines contribute will have smaller $P_{\rm N}$ value. Recently, \citet{Shaw2019} presented a methodology for incorporating the non-Gaussianity with the antenna baseline distributions and system noise to make more realistic error predictions for the future SKA-Low observations. They have also studied effectiveness of the inherent non-Gaussianity in presence of the system noise by comparing the Gaussian and the non-Gaussian error predictions. Figure \ref{fig:frac_1} shows the percentage deviation of the non-Gaussian error with respect to the Gaussian predictions for early (Top panel) and late (Bottom panel) stages of reionization for different $t_{\rm obs}$ values. Note that the deviation $\Delta$ is largest for the cosmic variance only case (CV) and it increases with $k$. However system noise contribution in variance prevails towards the larger $k$ modes as the baseline density decreases for larger $\vec{U}$ values. Hence the deviation $\Delta$ sharply decreases after the intermediate $k$ values in both the panels. It can also be noted that increasing $t_{\rm obs}$ decreases the system noise and the trispectrum contribution becomes more prominent. Note that, $\Delta$ is larger towards the later stages of reionization ($\xb =0.15$) for any particular $t_{\rm obs}$ as the non-Gaussian effects are stronger (CV lines) and also $P_{\rm N}$ decreases towards smaller redshifts. The impact of non-Gaussianity is significant at $k \lsim 0.1~{\rm Mpc}^{-1}$ and $k \lsim 1~{\rm Mpc}^{-1}$ during the early (Top panel) and late stages (Bottom panel), respectively. In a follow-up work, \citet{Shaw2020} have studied the effect of the non-Gaussianity on the errors in the inferred {\sc ReionYuga} model parameters.

For other statistics, such as MAPS \citep{Mondal2020} and bispectrum \citep{mondal21} one also expects similar qualitative behaviour of non-Gaussian effects in presence of system noise in their error covariances.


\subsection{Impact of first sources on statistical estimators}
\label{sec:source_impact}

Several studies have been performed so far to study the impact of the different kinds of first sources on the IGM via its impact on the different statistical measures of 21-cm signal from the CD-EoR. Depending on the length-scales of observations, the statistical measures such as variance, skewness, and power spectrum can distinguish `inside-out' and `outside-in' reionization scenarios \citep{watkinson14, majumdar16}. These statistics can also distinguish the isolated sources such as Pop III stars, galaxies, mini-QSOs, and HMXBs by measuring their signature on the 21-cm signal \citep{ghara16a,ross17,ross19,2019MNRAS.487.2785I}. It is possible to constrain these sources with high SNR using $\sim 1000$ hours of SKA-Low observations in future \citep{ghara16a}. The formation, growth and percolation of \HII~bubbles in the IGM during EoR carries imprints of ionizing sources. However the power spectrum, being a two-point statistics, is unable to capture these important non-Gaussian features in the EoR 21-cm signal. The power spectrum can react to the different source properties only up to a certain extent by showing the variation in its amplitude which arises due to the change in the Gaussian component of the signal.

Bispectrum is able to quantify some of the non-Gaussian imprints of the first sources in the EoR 21-cm field. It can respond to the signatures of different sources and processes more efficiently than the power spectrum by reflecting changes in both its sign and magnitude, as illustrated in Figure \ref{fig:PSBS_z} \citep{watkinson19,kamran21}. Further, the 21-cm bispectrum is also able to probe the IGM physics during CD by quantifying the non-Gaussianity invoked due to the different astrophysical processes \citep{kamran22}. Apart from these, the 21-cm bispectrum is also a robust way of distinguishing the dark matter models and can put a better constraint on the nature of the dark matter in the future EoR observations as compared to the power spectrum alone \citep{saxena20}. 


\subsection{Line-of-sight anisotropy effects}
\label{sec:los_effects}

The observations of the redshifted CD-EoR 21-cm signal along a particular line-of-sight (LoS) direction ($\nhat$) provide additional information related to the different unique local effects at the point of emission compared to the other cosmological signal. As already stated in section \ref{subsec:2.3.1}, the major LoS effects are the redshift space distortion (RSD) \citep{kaiser87,hamilton97}, light-cone (LC) \citep{matsubara97,barkana06}, and Alcock-Paczynski (AP) \citep{ap79} effects. The RSD comes into the picture due to the non-linear peculiar velocities of the gas particles that cause additional redshift or blueshift on top of the cosmological redshift, and hence distorts the signal along the LoS direction. The LC effect results from the redshift evolution of the signal along the LoS, as the 21-cm signal coming from different redshifts essentially belongs to different frequencies. The AP effect is quite different from these two effects as it is truly related to the geometry of the space-time, which is non-Euclidean.

These LoS effects imprint their signature on the statistical measures of the 21-cm signal by making the signal anisotropic along the LoS axis. In the view of the future SKA-Low interferometric observations of the CD-EoR, it is important to understand the impacts of these crucial effects on the signal for its accurate detection and correct interpretation. Therefore, one needs to consider these effects while estimating the CD-EoR parameters using various statistical estimates such as the power spectrum and bispectrum, etc. 

\citet{Bharadwaj_2004, Bharadwaj_2005}, for the first time, analytically pointed out the RSD effects in the context of the CD-EoR 21-cm signal. Their study showed that the peculiar velocities significantly change the magnitude and shape of the 21-cm power spectrum when computed using visibilities measured in a radio-interferometric observation. Using the inherent anisotropy in the 21-cm power spectrum, it is possible to extract the matter power spectrum during CD-EoR \citep[e.g.][]{barkana05, shapiro13}. \citet{majumdar13} are the first to properly implement RSD effects in the simulations of the EoR 21-cm signal. They have quantified the effect of RSD by estimating the monopole and quadrupole moments of the 3D power spectrum. The evolution in the quadrupole moment of the power spectrum with the mean neutral fraction $\xhb$ at large length-scales ($k \sim 0.12 ~{\rm Mpc}^{-1}$) can be used to track the reionization history to a great degree \citep{majumdar16}. \citet{ghara15a} and \citet{2021MNRAS.506.3717R} quantified the effect of RSD on 21-cm power spectrum during CD, when the fluctuations in the $T_{\rm s}$ controls the 21-cm fluctuations. They have reported that the effect of RSD on the CD 21-cm power spectrum is not too high as compared to its effect on EoR 21-cm power spectrum. 

Using a set of simulated 21-cm signal, \citet{datta2012} and \citet{datta14} have performed the first numerical investigation on the LC effect on the 21-cm power spectrum from the EoR. They have found that the LC effect significantly enhances the large-scale power spectrum and suppresses the small-scale power spectrum. They have also found that, during the EoR, the LC effect has the largest impact on the 21-cm power spectrum when reionization is $\sim 20\%$ and another when it is $\sim 80\%$ completed. However, they did not find that the LC effect introduces any significant LoS anisotropy to the power spectrum. \citet{ghara15b} made the first numerical study on the impact of the LC effect on the redshifted 21-cm power spectrum from CD. They have found that the impact of the LC effect is more dramatic when one considers the spin temperature fluctuations in the signal compared to the case when it is ignored.

The Alcock-Paczynski effect is another anisotropy in the signal. This makes any shape that is intrinsically spherical in nature appear elongated along the LoS due to the non-Euclidean geometry of the space-time. It is not much significant at low redshifts, but at high redshifts ($z \gtrsim 0.1$), this causes a substantial distortion in the signal, making the CD-EoR 21-cm power spectrum anisotropic along the LoS. \citet{ali05} was first to consider the AP effect in the context of the EoR 21-cm signal. They have quantified the relative contribution in anisotropy due to the AP effect when compared with the anisotropy due to the RSD and how they differ in their nature.

Apart from the power spectrum, the LoS anisotropy can affect the higher-order statistics too. \citet{majumdar20} and \citet{kamran21} for the first time quantified the impact of the RSD on the CD-EoR 21-cm bispectrum from the simulated 21-cm signal. They have found that depending on the length scales of the observation, RSD significantly impacts the magnitude, sign, and shape of the bispectrum. The LC effect also shows its significant impact on the 21-cm bispectrum during the EoR \citep{mondal21}. The RSD and LC effects are substantial enough and make them necessary to account for the correct interpretation of the signal statistics.

\subsection{Parameter estimation using models}
\label{sec:param_est}

We have seen in section \ref{sec:models} that the astrophysical information is embedded in the 21-cm signal through $x_{\rm HI}$ and $T_{\rm s}$ terms in the brightness temperature fluctuations (eq. \ref{eq:bt}). However the extraction of this information from a measurement of the signal is not trivial and an exploration of many theoretical models of the expected \HI~signal is necessary to interpret the measurements from radio observations. 

First of all, one should realize that it is not feasible to perform a full pixel-by-pixel comparison between the observed signal and theoretical models. Instead, the observed signal is first characterized by one or a combination of statistical estimators such as mean signal \citep[see e.g.,][]{singh2017, 2022JCAP...03..055G}, variance \citep[e.g,][]{patil2014}, power spectrum \citep[e.g,][]{Mertens2020,2021MNRAS.501....1G}, bispectrum \citep[e.g.,][]{kamran21}, bubble size distribution \citep[e.g.,][]{2018MNRAS.479.5596G}, Topological measurements \citep[e.g,][]{kapathia2021}, pattern recognition in 21-cm signal images \citep[e.g,][]{2019MNRAS.484..282G}, etc., to perform a comparison with theoretical models (see the discussion in section \ref{sec:stats}). Among these, the 3D power spectrum has been frequently used in many of the ongoing 21-cm observations. The preferred statistical measures often depend on instrumental sensitivity and might differ from the optimal methods that provide maximum information about the observed redshifts.

\begin{figure}[ht!]
\begin{center}
\includegraphics[width=0.5\textwidth]{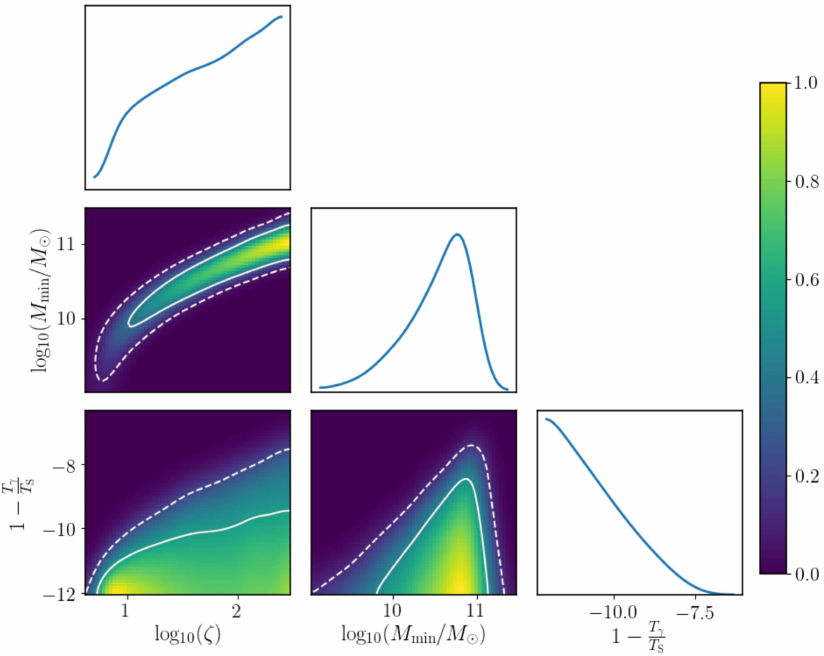}
\caption{Shows constraints on the {\sc grizzly} model parameters from the MCMC analysis using the LOFAR upper limit for $z \approx 9.1$. For this analysis, we assume uniform $T_{\rm s}$ for the neutral gas in the IGM. The colour bar represents the probability that models are ruled out by the upper limit. The $68\,(95)$ per cent credible intervals of the ruled out models are shown by the solid (dashed) curves. The marginalized probability distributions are shown in the diagonal panels. See \citet{2020MNRAS.493.4728G} for details.}
\label{fig:param_est}
\end{center}
\end{figure}

The next thing required for the parameter estimation is a fast method (see section \ref{sec:models}) that provides the above mentioned statistical quantities of the signal for a set of model parameters of our interest (e.g., the astrophysical parameters). For obtaining the probability distribution function (PDF) in the simulation model parameter space,  hundreds of thousands of models may be needed. In case running such a large number of models are computationally expensive, alternative approaches based on emulators \citep[e.g,][]{2020MNRAS.493.4728G}, neural networks \citep{2017MNRAS.468.3869S}, etc. might be taken for parameter inference. For example, \citep{2020MNRAS.493.4728G} build emulators based on Gaussian process regression which generate power spectrum for a set of source model parameters namely, ionization efficiency ($\zeta$), the minimum mass of the UV emitting halos ($M_{\rm min}$), the minimum mass of X-ray emitting halos ($M_{\rm min, X}$) and X-ray heating efficiency ($f_{\rm X}$).

Finally one needs to perform a robust probabilistic exploration of the source model parameters to unlock astrophysical information from the 21-cm measurements. To explore and constrain the reionization parameters, the signal prediction algorithms are often combined with a Bayesian inference framework. The likelihood used in the Bayesian analysis should take into account all possible uncertainties, both from observational and theoretical sides. It should also distinguish differences between an upper limit and a detection of the signal. \citep{2020MNRAS.493.4728G} combined {\sc grizzly} emulators with the Monte Carlo Markov Chains (MCMC) to constrain the source parameters at $z\approx9.1$ using the LOFAR upper limits on the 3D power spectrum of the 21-cm signal \citep{Mertens2020}. Figure \ref{fig:param_est} shows the constraints on the astrophysical parameters used in GRIZZLY as obtained from such an analysis. In addition to constraining the simulation astrophysical model parameters, one can also constraint the IGM quantities such as average ionization fraction, average gas temperature of the partially ionized IGM, that characterise size of the emission regions in the IGM, etc. \citep[see e.g.,][]{2020MNRAS.493.4728G, 2021MNRAS.503.4551G}.

\subsection{Image based statistics}

\begin{figure*}
	\centering
	\includegraphics[width=0.9\textwidth]{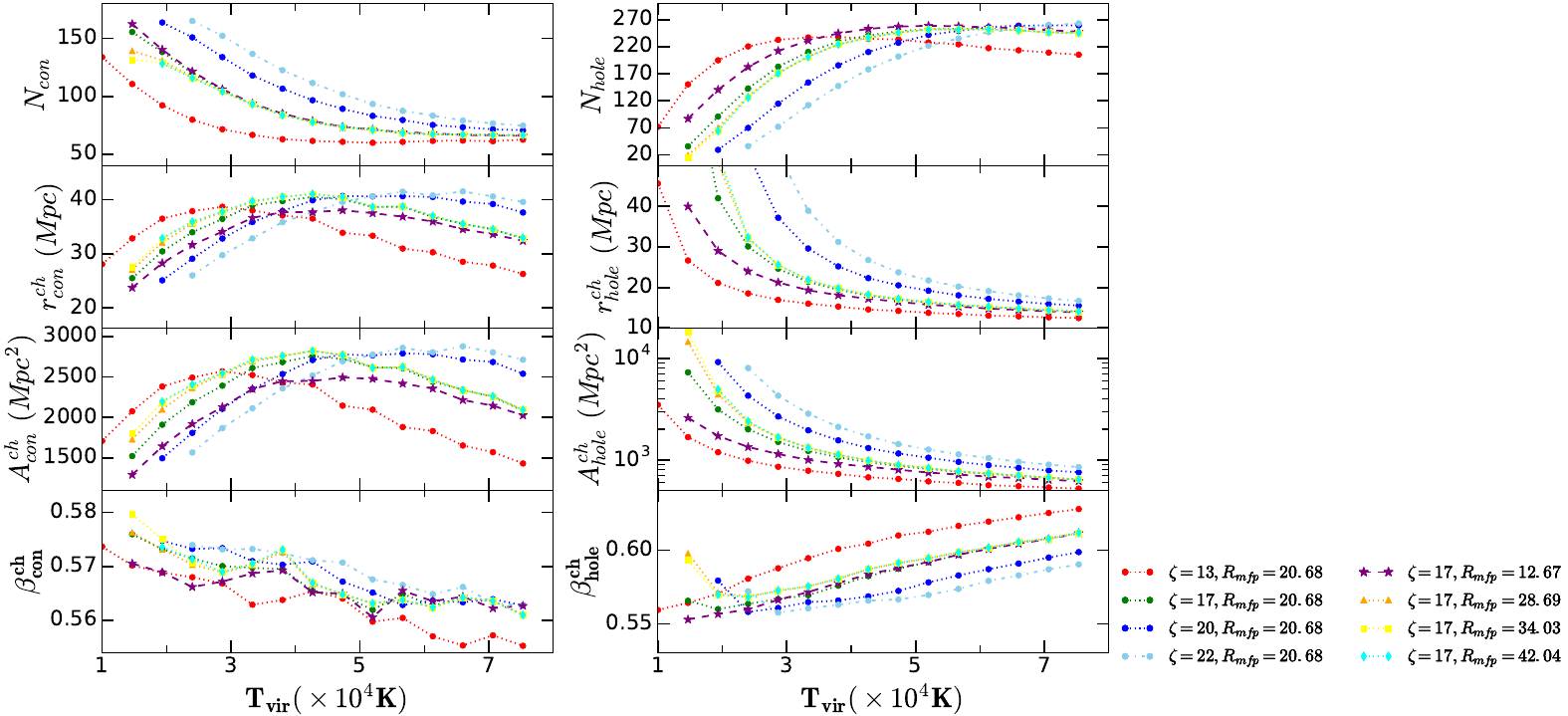}
	\caption{Shows the variation of hole (right panels) and connected region (Left panels) statistics with virial temperature $T_{\rm vir}$ for different fixed values of the ionization efficiency $\zeta$ and mean-free path $R_{\rm mfp}$. The simulations are done using {\sc 21cmFAST}. The top to bottom panels represent the threshold integrated number of district curves  $N_{\mathrm{x}}$, characteristic scale $r_{\mathrm{x}}^{ch}$, area $A_{\mathrm{x}}^{ch}$ and shape anisotropy parameter $\beta^{ch}_\mathrm{x}$ respectively. Suffix x=con stands for connected regions while x=hole refers holes. Credit: \citet{kapathia2021}.}
	\label{stat_fix}
\end{figure*}

Contrary to the currently operating radio interferometers which mainly focus to detect the signal statistically, the second phase of upcoming SKA-Low will be able to produce tomographic images of the CD-EoR 21-cm signal \citep{2015aska.confE..10M, ghara16}. Thus, the use of the CD-EoR tomographic images has been the focus of several theoretical studies which aim to develop methods for extracting information from such maps. Here we briefly mention some of these methods.

The expected 21-cm signal from the CD-EoR can be imagined as distributions of complex structures of ionized/neutral or hot/cold regions in space (as can be seen in Figure \ref{fig:HI_map}). Such complex topology is often studied in terms of the excursion set $F(\vec{x},u_{th})$ which represents the part of a given field with values larger than a certain field threshold $u_{th}$. Such excursion sets of the EoR 21-cm signal have been studied using four Minkowski functionals (MFs) to characterise the geometry of the 3D maps of reionization for different thresholds \citep[see e.g.,]{2006MNRAS.370.1329G, 2019ApJ...885...23C, FriedMF}. Among the four Minkowski functionals, three $V_{i=0,1,2}(u_{th})$ stands for the total volume, total surface and mean curvature of $F(\vec{x},u_th)$ respectively. The fourth functional $V_3(u_{th})$ is the integrated Gaussian curvature over the surface of $F(\vec{x},u_{th})$. Studies such as \citet{YoshiuraMF} show that these MFs are sensitive to the variation of reionization scenarios due to variation of the background source model. This shows the potential of using MFs in addition to the PS to tighten the constraints on the reionization parameters.

These MFs calculations can be used to characterise the shapes of compact surfaces. Shape-finder algorithms such as {\sc SURFGEN2} \citep[e.g.,][]{bag2019} use these MFs to define Thickness = $3V_0/V_1$, Breadth = $V_1/V_2$, Length = $V_3/4\pi$ to tract the evolution of the ionized/heated regions. While the Thickness and Breadth of the largest clustered region grow slowly during the entire reionization process, the Length evolves rapidly showing the percolation of the ionization regions into a filamentary structure \citep{bag2018}. Recently, \citet{pathak2022} studies the evolution of difference reionization scenarios using the Largest Cluster Statistics (LCS), which can be defined as the ratio of the volume of the largest ionized region and the total volume of all the ionized regions. This study shows  that  the percolation transition in the ionized regions, as probed by studying the evolution of the LCS, can robustly distinguish the inside-out and outside-in reionization scenarios.

Another generalization of the MFs is a rank-2 Contour Minkowski tensor which can provide alignment of 2D structures in $F(\vec{x},u_{th})$ and their anisotropy. The two eigenvalues of the Contour Minkowski Tensor $\lambda_1$ and $\lambda_2$ can be combined to construct morphological descriptors such as the shape anisotropy parameter as $\beta=\lambda_1/\lambda_2$, an effective radius of the enclosed curve $r = (\lambda_1+\lambda_2)/2\pi$. A region formed by a bounded curve enclosing a set of values larger than $u_{th}$ in the excursion set is known as a connected region. On the other hand, A hole represents a bound curve with values smaller than $u_{th}$. The count of these connected regions and holes are Betti numbers. \citet{kapathia2021} explore the prospects of constraining EoR scenarios using these morphological descriptors, Betti numbers, and the areas of structures in the connected regions and holes. The study shows that these morphological descriptors are sensitive to reionization parameters (see Figure \ref{stat_fix}) and can be used to constrain the source parameters as well as can put a strong bound on the ionization fraction at the probed reionization stage.

The complex structures in the ionization fraction maps and 21-cm maps of the EoR are often characterised in terms of distributions of spherical regions (often named as bubbles). Several studies using mean-free path \citep{2007ApJ...669..663M}, granulometry \citep{Koki2017}, Watershed methods \citep{Lin2016}, image segmentation method \citep{2018MNRAS.479.5596G} have aimed to study the complex morphology of the signal in terms of the size distribution of these bubbles and their evolution with time. As the size distribution of the bubbles vary significantly with change in reionization scenario, this in principle can be used for reionization parameter inference.

In addition to these, study of the tomographic images of the 21-cm signal using fractal dimension analysis can also provide distinguishable information between outside-in and inside-out reionization scenarios \citep[see e.g.,][]{2017MNRAS.466.2302B}. In addition, application of convolutional neural network constructed using 2D or 3D maps as training set is useful to distinguish between reionization scenarios and constrain astrophysical/cosmological parameters directly using a measured 21-cm input image \citep[see e.g.,][]{2019MNRAS.483.2524H, 2019MNRAS.484..282G}.


\section{Detection and Challenges}\label{sec:detection}
Detection of the 21-cm signal and extraction of the information content of it thereafter will bring to us a much clearer picture of the evolution of the cosmic matter since the epoch of  recombination. In this section we discuss different challenges identified in the detection of faint 21-cm signal along with some of the developments in literature that aim to resolve these problems. The majority of these problems come from the fact that the expected 21-cm signal is much fainter than other signals present in continuum in the same observing frequencies, termed collectively as foreground. Furthermore, the SKA Low and mid telescopes that would have expected sensitivity to measure the 21-cm signals would be able to only estimate statistical measure of the signals in its first phase. Keeping in mind the fact that these telescopes measure visibilities with incomplete baseline coverage, different unbiased estimators of the statistical properties of the signal are designed and  then tuned to address the problems of mitigating the unwanted signals in these frequencies. Recently, the importance of accurate calibration and the effect of uncalibrated part of the visibilities in detection of 21-cm signals are also being studied. We shall discuss these topics in this section. The section is organized such that the reader is informed about any challenge in the signal before the steps taken to find a resolution.

\subsection{Foregrounds}

We refer to  the radiation  from different astrophysical sources,  other than the cosmological HI signal,  collectively   as foregrounds. Foregrounds  include extragalactic point sources, diffuse synchrotron radiation from our Galaxy and low redshift galaxy clusters; free-free emission from our Galaxy (GFF) and external galaxies (EGFF). Extra-galactic point sources and the diffuse synchrotron radiation from our  Galaxy  largely dominate  the foreground radiation at $ 150\, \rm{MHz}$ and  their strength is $\sim  4-5$ orders of magnitude larger than the $\sim  20 -30 \,\rm mK$ cosmological 21-cm signal \citep{Ali_2008,Ghosh_2012}. The free-free emissions from our Galaxy and external galaxies make much smaller contributions though each of these is individually larger than the HI signal.

All the foreground components mentioned earlier  are continuum sources. It is well accepted that the frequency dependence of  the various continuum foreground  components can be modelled  by power laws \citep{Santos_2005},  and we  model the multi-frequency angular  power spectrum \citep{Kanan_2007} for each foreground component as

\begin{equation}
C_{\ell}(\nu_1,\nu_2)=A\left(\frac{1000}{\ell}\right)^{\beta}
\left(\frac{\nu_f}{\nu_1}\right)^{\alpha} 
\left(\frac{\nu_f}{\nu_2}\right)^{\alpha}
\label{eq:fg}
\end{equation}
where $A$ is the amplitude and   $\beta$  and $\alpha$ (spectral index) are the power law indices  for the $\ell$ and the $\nu$ dependence respectively. In general, we are interested in a situation where $\nu_2=\nu_1 + \Delta \nu$ with $\Delta \nu \ll \nu_1$, and we have  

\begin{equation}
C_{\ell}(\Delta \nu) \equiv C_{\ell}(\nu_1,\nu_1 + \Delta \nu) \approx A\left(\frac{1000}{\ell}\right)^{\beta}
\left(\frac{\nu_f}{\nu_1}\right)^{2 \alpha} \left(1 - \frac{\alpha \, \Delta \nu }{\nu_1}
\right)~,
\label{eq:fg1}
\end{equation}
which varies slowly with $\Delta \nu$. For the foregrounds, we expect  $C_{\ell}(\Delta \nu)$ to fall  by less than $10 \%$ if $\Delta \nu$ is varied  from $0$ to $3 \, {\rm MHz}$, in contrast to the $\sim 90 \%$ decline  predicted for the  HI  signal \citep{Bharadwaj_2005, ali2014}. The  frequency spectral index $\alpha$ is expected to have a scatter $\Delta \alpha$ in 
the range $0.1 -0.5$ for the different foreground components in  different directions  
causing  less than  $2 \%$  additional deviation in the frequency band of our interest. We  only refer the mean spectral indices
for the purpose of the foreground  presented here. In a nutshell, the   $\Delta \nu$  dependence  of $C_{\ell}(\Delta \nu)$ is  markedly different for the   foregrounds as compared to the  HI signal and would be very useful  to  separate  the foregrounds from the HI signal \citep{Ghosh_2012}. 

Extragalactic point sources are expected to dominate the 150 MHz sky at the angular scales relevant for redshifted 21 cm observations.
The contribution from extragalactic point sources is mostly due to the emission from normal galaxies, radio galaxies, star forming galaxies and active galactic nuclei \citep{Santos_2005, singal2010, condon2012}.

There are different radio surveys that have been  conducted at various frequencies 
ranging   from $151 \,{\rm MHz}$ to $8.5 \,{\rm GHz}$,  and these have a wide range of angular resolutions ranging  from $1^{''}$ to $5^{'}$ (eg. \citep{singal2010}, and  references therein). There is a clear consistency among the 
differential source count functions ($ \frac{dN}{dS} \propto S^{-\epsilon}$) at $1.4 \,{\rm GHz}$ for 
sources with flux $S > 1 \,{\rm mJy}$. The source counts are poorly constrained at $S \,< 1 \,{\rm mJy}$.  Based on the various radio observations \citep{singal2010}, we have identified  four different regimes  for the  $1.4 \,{\rm   GHz}$ source counts (a.)$\, \gsim 1\,{\rm Jy}$  which are the brightest
sources in the catalogs. These   are relatively nearby objects and they  have a  steep, Euclidean  source count with $\epsilon \sim 2.5 $;
(b.) $1\, {\rm mJy}$ - $ 1\, {\rm Jy}$ where the  observed differential source counts decline more gradually  with $ \,\epsilon \sim 1.7
$ which  is caused by redshift effects; (c.) $ 15 \, \mu {\rm Jy}$ - $\, 1\,
{\rm mJy}$ where the source counts are again steeper  with $\epsilon > 2$  which is  closer to Euclidean,  and there is considerable scatter from field to field;   and (d.) $\lsim\, 15 \,\mu {\rm Jy}$,  the source counts must eventually flatten 
($\epsilon < 2$) at  low $S$  to avoid an infinite integrated flux. The 
cut-off  lower flux where the power law index $\epsilon$ falls  below $2$ is not well established, and  deeper radio observations are required. The first turnover or flattening flux in the  $ 1.4\, \rm GHz$ differential source count has been reported  at $\sim \,1\,\rm mJy$ \citep{condon1989, hopkins2003, owen2008} and it  is equivalent to $\sim 5 \,\rm mJy$ at $150\, {\rm MHz}$ assuming a spectral index of 0.7 \citep{blake2004, randall2012}.

The analysis of large samples of nearby radio-galaxies has shown that the point sources are clustered \citep{cress1996,wilman2003, blake2004}.  The measured two point correlation function can be well fitted with a single power law  $w({\theta})=(\theta/\theta_{0})^{-\beta}$ where $\theta_{0}$ is the correlation length. We can calculate  $w_{\ell}$ which is $\propto {\ell}^{\beta -2}$ and  the Legendre transform of $w(\theta)$. Using this and source counts we have modeled the angular power spectrum due to the clustering  and Poisson contribution of point sources \citep{Ali_2008, ali2014}.

The galactic diffuse synchrotron radiation is believed to be produced by cosmic
ray electrons propagating in the magnetic field of the galaxy \citep{rybicki1979}. Angular structure in the
diffuse Galactic emission has been shown to be well described by a power-law spectrum in Fourier space over a large range
of scales. \citet{laporta2008} have determined  the angular power spectra
of the Galactic synchrotron emission at angular scales greater 
than $0.5^{\circ}$  using total intensity all sky maps at $ 408 \,\rm MHz $ \citep{haslam1982}  and $1.42 \, \rm GHz$ \citep{reich1982, reich1986}. However, there are only a
few observations which have directly measured $C_{\ell}$ at the frequencies
and angular scales which are relevant to the EoR studies \citep{Bern09,parsons2010,Ghosh_2012,iaco13,Samir_2020}. They find that the 
 angular power spectrum of synchrotron emission is well described 
by a power law   (eq. \ref{eq:fg}) where the value  of $\beta$ varies  in the range $1.8$ to $3.0$ depending on the 
galactic latitude.  \citet{laporta2008} have analyzed the frequency dependence  to find $A \propto \nu^{-2 \alpha}$ with $\alpha$ varying in the range  $2.8$ to $3.2$. There is a modest variations in the spectral index as a function of direction and frequency. The mean spectral index of the synchrotron emission at high Galactic  latitude has been recently  constrained to be $\alpha \approx 2.5$ in
the $150 - 408 \, \rm MHz$ frequency range \citep{rogers2008} using  single dish observations. In general, it is steeper at high Galactic latitudes than toward the Galactic plane. The Galactic Free–Free (GFF) and the Extra-Galactic Free–Free (EGFF) components
which are relatively much weaker foregrounds as compared to earlier two. But both are stronger than the  redshifted 21 cm signal. Their $\alpha$ and $\beta$ are presented  at $130  \, \rm MHz$ \citep{Santos_2005}.


\subsection{Statistical detection of the 21cm signal}\label{subsec:stat-detect}
Most of the radio interferometers measure visibility function sampled at certain baseline positions. Here we discuss development of various visibility based estimators of the power spectrum of the sky signal. These estimators are unbiased within the scope of the signal considered. Different estimators discussed here also address and solve various challenges that one face for 21-cm detection.

\subsubsection{Bare Estimator}
The Bare estimator measures the angular power spectrum $(C_{\ell})$, which quantifies the intensity fluctuations in the two-dimensional sky plane. It uses individual visibilities to measure $C_{\ell}$. As the visibility $\V_i$ at a baseline $\u_i$ corresponds to a Fourier mode in the sky, the two visibility correlation straight away gives the angular power spectrum which can be written as,

\begin{equation}
\langle \V_i \V^{*}_j \rangle = V_0 \, e^{-\mid \Delta \u_{ij} \mid^2/
\sigma_0^2} 
\, C_{\ell_i} + \delta_{ij} 2 \sigma_n^2
\label{eq:vcorr}
\end{equation}
where $V_0= \frac{\pi \theta_0^2}{2}
\left( \frac{\partial B}{\partial T}\right)^{2}$,  $\Delta \u_{ij}=
\u_{i}-\u_{j}$ and the Kronecker delta $\delta_{ij}$ is nonzero only if 
we correlate a visibility with itself. For the Gaussian approximation of the primary beam $\theta_0=0.6 \theta_{\rm FWHM}$, $B$ is the Planck function  and  $({\partial B}/{\partial T})=2
k_B/\lambda^2$ in the Raleigh-Jeans limit which is valid at the frequencies considered here. The $\sigma_n$ is the rms of the real and imaginary part of noise in the measured visibilities. We thus see that the visibilities at two different
baselines $\u_i$ and $\u_j$ are correlated only if the separation
is small $(\mid \Delta U \mid \le \sigma_0)$, and  correlation falls as the separation is beyond a disk of radius $\sigma_0$.

To avoid the positive noise bias in the second term of eq. \ref{eq:vcorr}, we define the Bare\footnote{\url{https://github.com/samirchoudhuri/BareEstimator}} estimator as 

\begin{equation}
\hat E_B(a)=\frac{\sum_{i,j} \, w_{ij} \, \V_{i} \,  \V^{*}_{j} }{\sum_{i,j} w_{ij} V_0  
e^{-\mid \Delta \u_{ij} \mid^2/\sigma_0^2} } = \frac{Tr({\bf w} {\bf V}_2)}{Tr({\bf w} {\bf I}_2)}.
\label{eq:be1}
\end{equation}

 The weight 
$w_{ij}=(1-\delta_{ij})$ is chosen such that 
it is zero when we correlate a visibility with itself, thereby avoiding the positive noise bias. We have the matrices ${\bf w} \equiv w_{ij}$, ${\bf V}_2 \equiv \V_{i} \,  \V^{*}_{j}$, ${\bf I}_2= V_0  
e^{-\mid \Delta \u_{ij} \mid^2/\sigma_0^2}$ and $Tr({\bf A})$ denotes the trace of a matrix ${\bf A}$.

The variance of the Bare estimator can be simplified to 
\begin{equation}
\sigma^2_{E_B}(a)=\frac{\sum_{i,j,k,l} w_{ij} w_{kl} V_{2il} V_{2kj}}{[Tr({\bf w} 
{\bf I}_2)]^2}
=\frac{Tr({\bf w} {\bf V}_2 {\bf w} {\bf V}_2)}{[Tr({\bf w} {\bf I}_2)]^2}
\label{eq:be6}
\end{equation} 
under the assumptions  that  ${\bf w}$ is symmetric. The detailed formalism of the Bare estimator and the validation with realistic simulations are given in Section 4 of \citep{Samir_2014}.

\subsubsection{Tapered Gridded Estimator}

Point sources are the most dominant foreground components at angular scale $\le4^{\circ}$ \citep{Ali_2008}. Due to the highly frequency-dependent  primary beam, it is very difficult to remove the point sources from the outer edge of the primary beam. These outer point sources create a oscillation along the frequency axis \citep{Ghosh_2011} and make it difficult to remove under the assumption of the smoothness along with the frequency.  This issue is not addressed in the bare estimator.  The  Tapered Gridded Estimator  (TGE\footnote{\url{https://github.com/samirchoudhuri/TGE}})  incorporates three novel features: First, the estimator uses the
gridded visibilities which makes it computationally much faster for large data volume. Second, the noise bias is removed by subtracting the auto-correlation of the visibilities from each grid point. Third, the estimator also taper the FoV to restrict the contribution from the sources in the outer regions and the sidelobes. The mathematical formalism and the variance of the TGE are given in eq. (17) and eq. (25) of \citep{Samir_2016}. 

\begin{figure}[ht!]
	\centering
	\includegraphics[width=.5\textwidth]{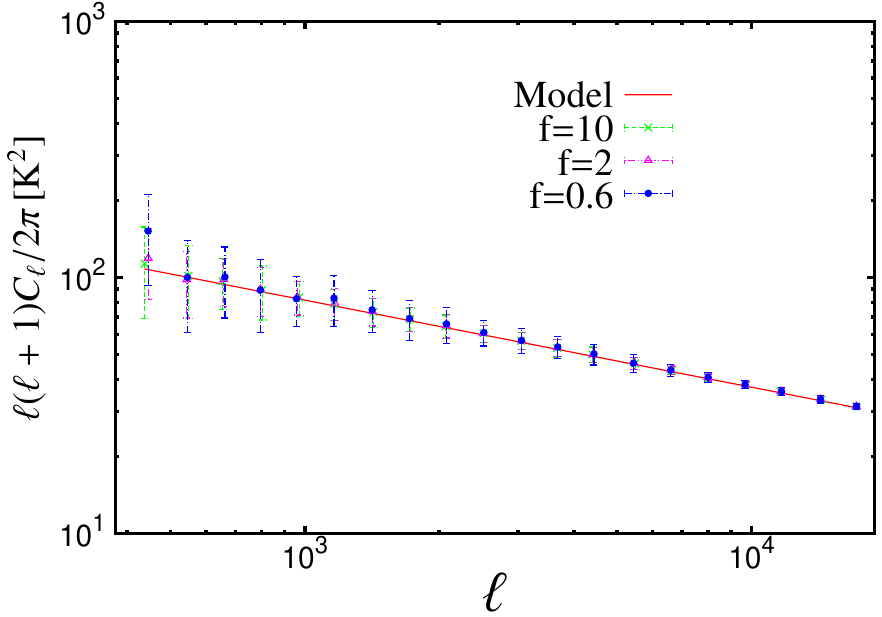}
	\caption{Shows the estimated values of $C_{\ell}$ from the simulated visibilities using the TGE for
		different tapering of values $f=10,2$ and $0.6$, with 1-$\sigma$ error bars
		estimated from $N_r=128$ realizations of the simulations. We also show the input model with red solid line for comparioson. This figure is adapted from Figure2 of \citep{Samir_2016}.}
	\label{fig:tge2d}
\end{figure}

In Figure \ref{fig:tge2d} we show the validation of the TGE using realistic simulation of GMRT observations at 150 ${\rm MHz}$ . The red solid line shows the model power-law power spectrum $C^M_{\ell}=A_{\rm 150}\left(1000/\ell \right)^{\beta} $used for the simulation. Here, we have used  $A_{\rm 150}=513 \, {\rm  mK}^2$ and  $\beta=2.34 $. The different color points are for different types of the tapering values $f=10, 2$ and $0.6$. The value of $f=10$ essentially corresponds to a situation with no
tapering, and the sky  response gets confined to a progressively smaller 
region  as the value of $f$ is reduced to $f=2.0$ and $0.6$ respectively. The points here indicates the estimated values of the $C_{\ell}$ with  $1\sigma$ error
bar after applying the TGE. We see that the TGE is able to recover the
input model $C^M_{\ell}$ quite accurately for all  values of tapering function $f$. 

\subsubsection{Image based Tapered Gridded estimator}
\label{sec:itge}
The effect of sources outside the field of view of an interferometric observations for the estimation of the power spectrum of redshifted 21-cm emission is addressed in TGE by introducing an analytical tapering function in the visibility plane \citep{Samir_2016}. In such observations of 21-cm signal  significant contribution from localized  foreground emission from within the field of view also can overwhelms the \HI\ signal. These localized emission are often compact and can be subtracted from the visibilities with a reasonable model by the method of uvsub. However,  it is rather difficult to sufficiently accurately model localized but extended sources and a simple uvsub may not work. Addressing this issue with the TGE is also difficult owing to the complicated nature of window that one need to model analytically to reduce the foreground effects. It is possible to introduce an image based tapering algorithm, where a wide variety of non trivial and non analytical tapering functions  can be used. This method of estimating the power spectrum directly from visibilities with arbitrary tapering window implementation in the image plane is implemented in the  Image based Tapered Gridded Estimator (henceforth ITGE). A detailed discussion of the algorithm along with  diagnostics tests are discussed in \cite{samir-19}, we give a brief description here.

\begin{figure}[ht!]
	\begin{center}
		\includegraphics[width=0.48\textwidth]{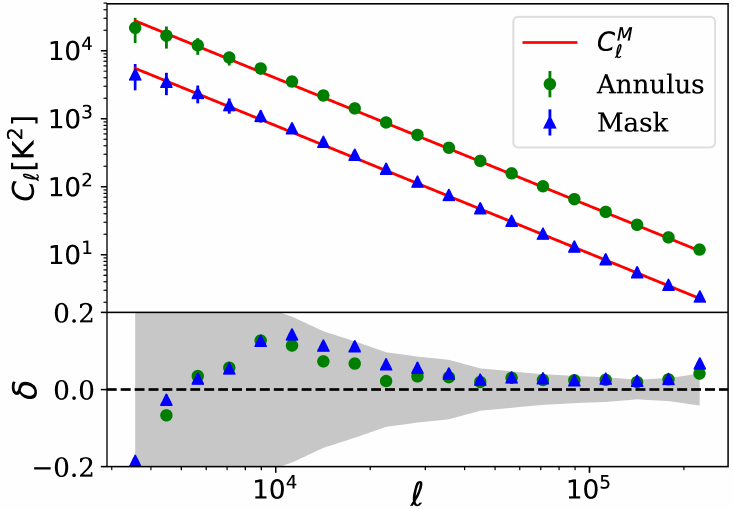}
		\caption{This figure shows the estimated and model $C_{\ell}$ using ITGE (section~\ref{sec:itge}) for two different type of windows applied in the image plane: (a) 2D Annulus window and (b) Mask window with three holes present within the mainlobe of the primary beam. We also show the model $C^M_{\ell}$ with red solid lines. The lower panel shows the fractional deviation between the estimated and model angular power spectrum. The shaded region in the lower panels show the expected statistical fluctuations $\sigma/C^M_{\ell}$. For more details we refer the reader to Figure4 of \citep{samir-19}.}
		\label{fig:cl_ITGE}
	\end{center}
\end{figure}

As for the case of TGE, the basic expression that coins the estimation of power spectrum at a grid is given in eq. (17) of \citep{Samir_2016}. To use an image based tapering window the following steps are followed. First the visibilities are gridded in baseline plane with suitable choice of grid size and weighting schemes. A dirty image is made from the gridded visibilities and multiplied with the chosen tapering window $W({\theta})$ in the image plane. An inverse Fourier transform of this provides the tapered and convolved visibilities in a grid. Apparently, these gridded visibilities then can be squared to estimate the power spectrum. However, as discussed for the TGE, this would introduce a large bias from the noise contribution in the self visibility correlations. To avoid the  noise bias in the estimated power spectrum, the TGE algorithm need to estimate the  self correlations of  tapered visibilities in each grid. In ITGE algorithm, the correlations of the visibilities in each grids are estimated. In parallel to this, the Fourier transform of the tapering window $\tilde{w}_g$ is evaluated at the grids of the baseline plane. Two dirty images, one corresponds to the visibility self correlation and other of $\mid \tilde{w}(\vec{U}) \mid^{2}$ are estimated and multiplied in the image plane. Inverse Fourier transform of the later gives the self correlations of  tapered visibilities in each baseline grid. Rest of the algorithm as well as the error estimation follows directly from the original TGE. 

\citet{samir-19} have also validated the estimator for two window functions namely (a) the Annulus window, and (b) the Mask
window. The Annulus window allows us to estimate $C_{\ell}$ only from the outer annular region, where as the Mask window has three holes within the mainlobe of the primary beam. The upper panel of Figure \ref{fig:cl_ITGE} shows the estimated $C_{\ell}$ along with the input model $C_{\ell}$ for these two different windows applied in the image plane. The results have been arbitrarily scaled for clarity of
presentation. The lower panel of Figure \ref{fig:cl_ITGE} shows the fractional deviation between the estimated and model angular power spectrum. The shaded region in the lower panels show the expected statistical fluctuations $\sigma/C^M_{\ell}$. We see that the fractional deviation is less than $15 \%$ for whole $\ell$ range, also within $1~\sigma$ of the statistical fluctuations.

\subsubsection{3D Tapered Gridded Estimator}
The power spectrum estimator addressed so far estimates the power spectrum on the sky plane at the same frequency channel. To measure the fluctuations along the line of sight or, equivalently, along frequency directions, we have further extended the two-dimensional TGE to three-dimensional TGE which measures the fluctuations from a 3D cube. Here we first multiply the data along the frequency with a Blackman-Nuttall \citep{Nuttall81} window function to make data periodic, and then use Fastest Fourier Transform in the West (FFTW) \citep{Frigo_05} algorithm along the frequency axis to convert the visibilities in delay space $(\tau)$, the Fourier conjugate of the frequency $(\nu)$. After conversion in delay space, we apply the same 2D TGE (earlier section) in each slice of $\tau$, and further, we spherically bin the data to measure the power spectrum as a function of $k$. The mathematical expression for the 3D TGE (eq. 44 of \citep{Samir_2016} is

\begin{equation}
\begin{split}
{\hat P}_g(\tau_m)= & \left(\frac{M_g {\rm B_{bw}}}{r^2 r^{'}} \right) ^{-1} \,
( \mid v_{cg}(\tau_m) \mid^2 - \\
& \sum_i \mid \tilde{w}(\u_g-\u_i) \mid^2  
\mid v_i(\tau_m) \mid^2 ) \,. 
\end{split}
\label{eq:a16}
\end{equation}

where $v_i(\tau_m)$ and $v_{cg}(\tau_m)$ are the individual and convolved-gridded visibilities in delay $(\tau)$ space respectively, $\tilde{w}(\u)$ is the Fourier transform of the tapering window function used to suppress the primary beam sidelobes in the sky plane, $M_g$ is the normalization constant and it depends on the baseline distribution and the form of the tapering function. We used simulated visibilities corresponding to an unit angular power spectrum (UAPS) $C_{\ell}=1$ to estimate $M_g$. The ${\rm B_{bw}}$ is the bandwidth of the observation,  $r$ is the comoving distance corresponding to the redshifted 21-cm radiation at the observing frequency $\nu$, $r^{'}=\mid dr/d \nu \mid$.

\begin{figure}[ht!]
\begin{center}
\includegraphics[width=0.48\textwidth]{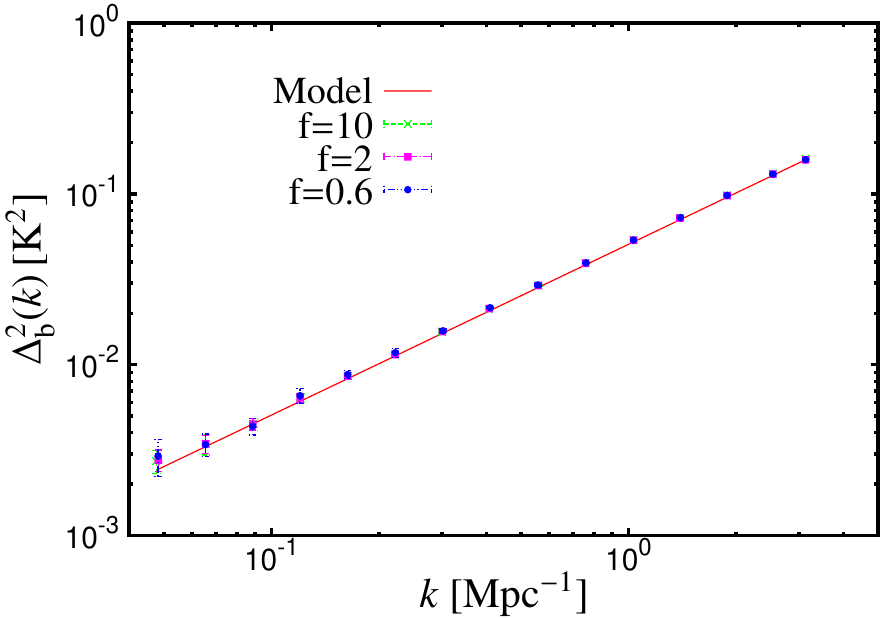}
\caption{Shows the dimensionless power spectrum $\Delta_{\rm b}^2(k)$ for different values of tapering function $f$. The values obtained using the 3D TGE are compared with model power spectrum shown with red solid line. The 1-$\sigma$ error bars are estimated using 16 different realizations of the simulated visibilities. This figure is adapted from Figure7 of \citep{Samir_2016}.}
\label{fig:tge3d}
\end{center}
\end{figure}

Figure \ref{fig:tge3d} shows the dimensionless spherically-averaged power spectrum for the model $P(k)=k^{-3}$. The results are shown for the three different values of tapering windows parametrized by $f=10 ,2$ and $0.6$ to demonstrate the effect of varying the tapering. For all the values of $f$ we find that the estimated power spectrum using the 3D TGE is within the $1-\sigma$ error bars of the model prediction for the entire $k$ range considered here.

\subsubsection{The MAPS and The Power Spectrum}
Nearly all the estimators for $P(k_{\perp},k_{\parallel})$ \citep{morales05,parsons12,liu14a,liu14b,trott16,Mertens2020}, including the 3D-TGE (see earlier subsection), consider a Fourier transform (FT) of the measured visibilities $\V_i(\nu)$ along the frequency $\nu$ to obtain the visibilities $\V_i(\tau)$ in the delay space $\tau$, which is then used to estimate the cylindrical power spectrum $P(k_{\perp},k_{\parallel})$. In a real observation, particularly at low frequencies, quite often many of the frequency channels are flagged. As the quantity  $\V_i(\tau)$ is estimated by a direct Fourier transform of the frequency axes of the visibilities, this gives rise to  a convolution with the FT of the frequency sampling function, and thus introduces artifacts in the measured $P(k_{\perp},k_{\parallel})$.  To mitigate this problem, \citet{Somnath19} further developed the MAPS based TGE estimator that estimates the power spectrum of the cosmological 21-cm signal by first estimating the MAPS $C_{\ell}(\Delta\nu)$ and subsequently by taking a FT of the $C_{\ell}(\Delta\nu)$ along the $\Delta \nu$. Note that the MAPS based TGE estimator preserves all the salient features of the 2D and 3D TGE mentioned in the earlier subsection, while also providing a way to eliminate the challenges presented by the flagging usually prevalent in the observed radio data.

The details of the MAPS based estimator can be found in \citet{Somnath19} and \citet{Pal21}. To summarize, we first estimate the gridded visibilities $\V_{cg}(\nu_a)$ at a grid point $\u_g$ on a rectangular grid in the uv plane, by convolving the visibilities $\V_i(\nu_a)$ with the window function $\tilde{w}(\u)$ at all the frequency channels $\nu_a$ where $1 \le a \le N_c$, $N_c$ being the total number of channels. The convolution of the visibilities in the uv plane modulates the PB pattern by a multiplication in the image domain with the FT of the convolving function $\tilde{w}(\u)$, which is suitably chosen to taper the sky response away from phase center and especially outside the main lobe of the PB.
We then have,
\begin{equation}
\V_{cg}(\nu_a) = \sum_{i}\tilde{w}(\u_g-\u_i) \, \V_i(\nu_a) \,F_i(\nu_a),
\label{eq:cltge1}
\end{equation}
where, the variable $F_i(\nu_a)$ incorporates the flagging, and has a value `$0$' for the flagged data and `$1$' otherwise.
The TGE for  $\cl(\nu_a,\nu_b)$ is then defined as,
\begin{eqnarray}
&&{\hat E}_g(\nu_a,\nu_b) = M_g^{-1}(\nu_a,\nu_b) \, 
{\mathcal Re} \Big[\V_{cg}(\nu_a) \,  \V_{cg}^{*}(\nu_b) \, \nonumber \\
&& - \, \sum_i F_i(\nu_a)F_i(\nu_b) \mid
\tilde{w}(\u_g-\u_i) \mid^2   \V_i(\nu_a) \V_i^{*}(\nu_b)  \Big] \, \nonumber \\
\label{eq:cltge2}
\end{eqnarray}
where ${\mathcal Re}[..]$ denotes the real part of the terms within $[..]$ and $M_g(\nu_a,\nu_b)$ are the normalization factors determined using simulations. Note that the second term in the brackets denotes the self-correlation of the visibilities, and are subtracted out to eliminate the noise bias from the estimator considering that the noise in the visibility measurements at different baselines and timestamps are uncorrelated. The TGE described in eq.~(\ref{eq:cltge2}) provides an unbiased estimate of the $\cl{_g}(\nu_a,\nu_b)$ at the angular multipole $\ell_g=2 \pi U_g$, {\it i.e.},
\begin{equation}
\langle {\hat E}_g(\nu_a,\nu_b) \rangle = \cl{_g}(\nu_a,\nu_b)
\label{eq:cltge3}
\end{equation}

The MAPS $C_{\ell}(\nu_a,\nu_b)$ does not assume any ergodicity of the sky signal along the line of sight and can be used to quantify the second order statistics of the cosmological 21-cm signal including the light-cone effect for wide-bandwidth observations. Conversely, considering a small observation bandwidth, the redshifted 21-cm signal can be assumed to be statistically homogeneous (ergodic) along the line of sight (e.g. \citealt{Mondal2018}), allowing us to express $\cl(\nu_a,\nu_b)$ (eq.~(\ref{eq:cltge2}),~(\ref{eq:cltge3})) in terms of $\cl(\Delta \nu)$ where $\Delta \nu=\mid \nu_b-\nu_a\mid$. This transformation implies that the statistical properties of the signal depends only on the frequency separation and not the individual frequencies. In the flat sky approximation, the power spectrum $P(k_{\perp},k_{\parallel})$ of the brightness temperature fluctuations of the redshifted 21-cm  signal is then given by \citep{Kanan_2007},
\begin{equation}
P(k_{\perp},\,k_{\parallel})= r^2\,r^{\prime} \int_{-\infty}^{\infty}  d (\Delta \nu) \,
  e^{-i  k_{\parallel} r^{\prime} \Delta  \nu}\, \cl(\Delta \nu)
\label{eq:cl_Pk}
\end{equation}
where $k_{\parallel}$ and $k_{\perp}=\ell/r$ are the components of $\bf{k}$ respectively parallel and perpendicular to the
line of sight, $r$ is the comoving distance and $r^{\prime}~(=d r/d \nu)$, both being evaluated at the central observation frequency $\nu_c$. Starting from eq.~(\ref{eq:cl_Pk}), \citet{Pal21} have shown that given the estimates of $C_{\ell}(n\Delta\nu_{c})$, $\Delta\nu_{c}$ being the smallest channel separation with $0\le n \le N_{c}-1$, the maximum likelihood estimate (MLE) of the Cylindrical power spectrum $\bar{P}(k_{\perp},k_{\parallel m})$ is  given by,
\begin{equation}
\bar{P}(k_{\perp},k_{\parallel m}) =  \sum_n \{ [\textbf{A} ^{\dagger} \textbf{N}^{-1} \textbf{A}]^{-1} \textbf{A}^{\dagger}    \textbf{N}^{-1} \}_{mn}  C_{\ell}(n\Delta\nu_{c})
\label{eq:ML}
\end{equation}
where \textbf{A} refers to the $N_c \times N_c$ Hermitian matrix corresponding to the FT coeﬃcients, { \bf N} refers to the noise covariance matrix and `$\dagger$' denotes  the Hermitian conjugate.

\begin{figure}[ht!]
	\begin{center}
		\includegraphics[width=0.48\textwidth]{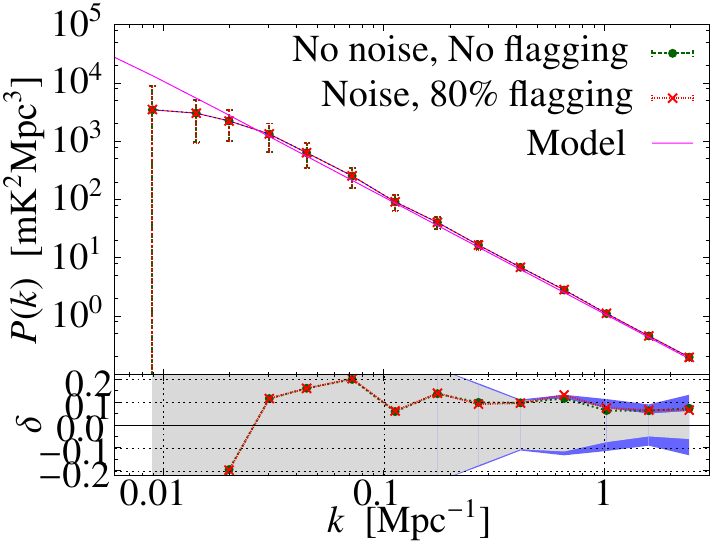}
		\caption{The upper panel shows the estimated spherically-binned power spectrum $P(k)$ along with $1-\sigma$ error-bars obtained from the simulations with no noise and flagging and also with noise and $80 \%$ flagging. The input model $P^{m}(k)$ is shown by the solid magenta line for comparison. The bottom panel shows the fractional error $\delta=[P(k)-P^m(k)]/P^m(k)$  (shown by the data points)  and the relative statistical fluctuation $\sigma/P^m(k)$ (shown by the shaded regions). See \citet{Somnath19} for details.}
		\label{fig:PK_MAPS_TGE}
	\end{center}
\end{figure}

The MAPS based TGE estimator have been validated using realistic simulations of GMRT observations at $150\,\,{\rm MHz}$. Using a power-law model power spectrum $P^{M}(k)$, \citet{Somnath19} have shown that the TGE can recover the input MAPS as well as the $P^{M}(k)$ with an accuracy of $5-20\%$ over a large $k$ range, even in the presence of the system noise and when $80\%$ of the data, chosen randomly, are flagged (see Figure~\ref{fig:PK_MAPS_TGE}). The signal in the visibility measurements $\V_i(\nu_a)$ at different baselines $\u_i$ are not independent, and gets correlated at baselines within a scale of $D/\lambda$ due to the telescope’s PB pattern, $D$ being the antenna diameter. Similarly, the visibility measurements $\V_i(\nu_a)$ at different $\nu_a$ are not independent and the signal gets correlated across different channels, the width of the correlation depending on the value of $\u_i$ (See \citealt{Bharadwaj-Pandey-2003,Bharadwaj_2005} for details). The signal contained in the flagged data which is lost due to the flagging thus also gets contained in the valid data which is available to us for the power spectrum estimation, and consequently the MAPS based TGE is able to recover the power spectrum equally well even if $80\%$ of the data is flagged.

\subsection{Observations with the GMRT}

\begin{figure*}[ht!]
	\centering
	\includegraphics[width=0.9\textwidth,angle=0]{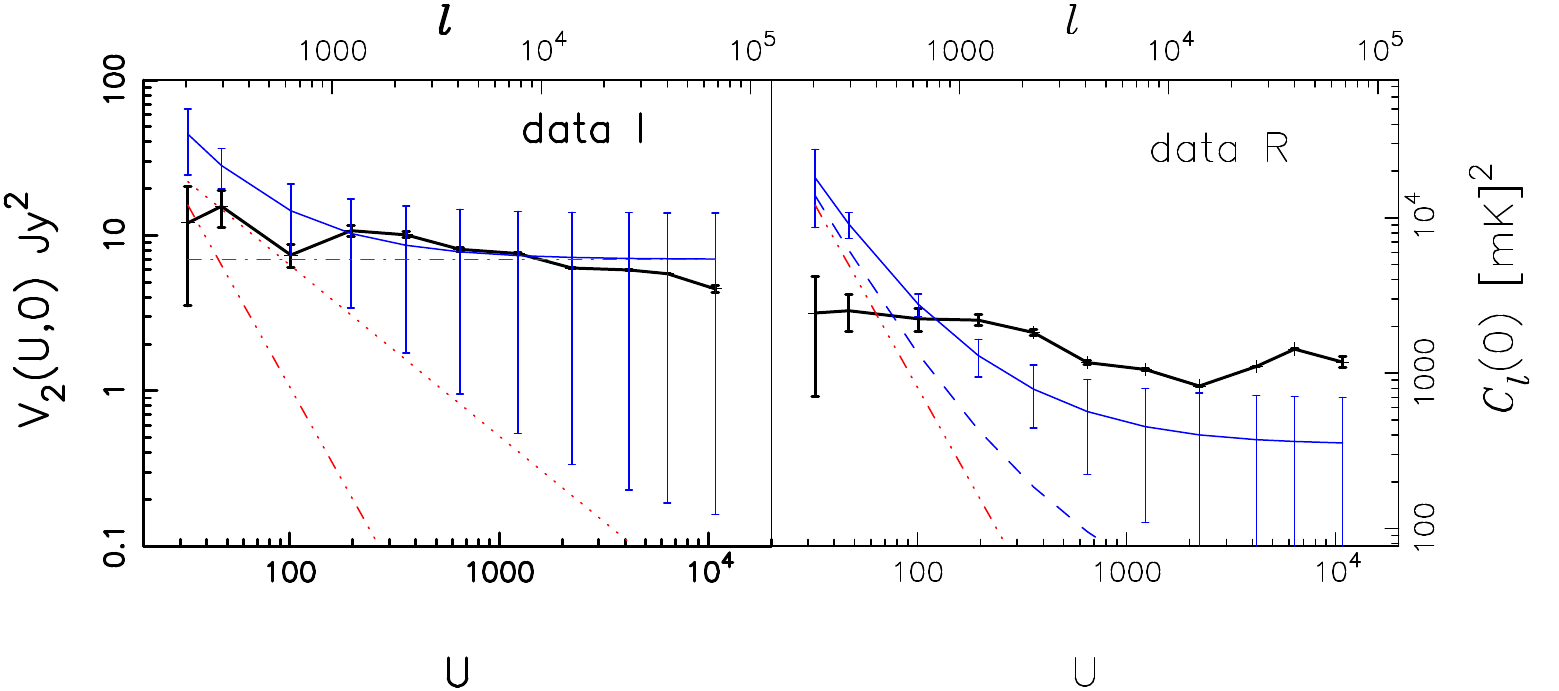} 
	
	\caption{Shows the visibility correlation as a function a function of baseline distance for total data set ($I$; left panel) and the residual data set ($R$; right panel). For data $I$, the thin solid line shows the total model prediction for $S_{c}$ = $900$ mJy. Also shown are the contributions from point source Poisson distribution (dash–doted), point source clustering (dot) and Galactic synchrotron (dash–dot–dot–dot). For data R, the thin solid line shows the total model predictions for $S_{c}$ = $100$ mJy and the long dashed line for 10 mJy. The dash–dot–dot–dot curve shows the Galactic synchrotron contribution. } 
	\label{fig:Saiyad_2008}
\end{figure*}

The Giant Metrewave Radio Telescope (GMRT) is one of  the  largest  and  most  sensitive  fully  operational  low-frequency  radio  telescopes  in  the  world  today  \citep{Swarup_1991} with $30$ fully steerable $45$ m diameter dishes spread with maximum antenna separation of $25$ km. Recently GMRT has been upgraded to uGMRT with some extra features, such as: (i) huge frequency coverage, from 120-1500 MHz; (ii) maximum bandwidth available is $400$ MHz instead of $32$ MHz bandwidth of original GMRT design; (iii) digital backend correlator catering to $400$ MHz bandwidth; (iv) improved  receiver  systems  with higher $G/T_{sys}$ and  better  dynamic  range \citep{Yashwant_2017}. The uGMRT is considered as one of the pathfinder of the SKA telescopes.

The expected brightness temperature of redshifted the \HI~signal from the EoR and post-reionization epoch  is many orders of magnitude fainter than the radio emissions from  different Galactic and extragalactic foregrounds \citep{Zaldarriaga_2004,Bharadwaj_2005,Santos_2005}. The challenges are nearly identical for both EoR and post-reionization experiments. So the knowledge of  foregrounds at post-reionization epoch can also help us to understand the intricacies involved in  detection of the \HI~signal coming from EoR. Accuracy of  extraction of the cosmological \HI~signal strongly depends on the ability to characterize and remove the foregrounds from observational data sets at the frequency of redshifted \HI~21-cm line. There have been a considerable amount of efforts put forward over the last decade in  characterizing the statistical nature of foregrounds, calibration issues, instrumental systematics and putting upper limits on \HI\ signal both from EoR and post-EoR epochs using sensitive low-frequency observations with GMRT. Here we will briefly mention different findings using low-frequency radio observations so far with GMRT.

\citet{Ali_2008} observed a field centred on Upsilon Andromedae, an extra-solar planetary system, at $\alpha_{2000}=1^{h}36^{m}48^{s} ,\delta_{2000}=41^{\circ}24'23''$  for a total of 14 hours using  GMRT Software Backend (GSB) system. The observation was conducted at 153 MHz, corresponding to the redshift $z \sim 8.28$, with 8 MHz bandwidth sub-divided into 128 frequency channels, giving a spectral resolution of 62.5 KHz. They have studied, for the first time,  the statistical nature of foregrounds on sub-degree angular scales using this observation at a redshift corresponding to the epoch of reionization. They have used the correlations among  measured visibilities to directly determine the multi-frequency angular power spectrum $\cl$ $(\Delta \nu)$ (MAPS, \citet{Kanan_2007}). The visibility correlation as a function of baseline distance or the angular scale $l$ is shown in Figure \ref{fig:Saiyad_2008}. They have found that the measured $\mathcal{C}_{\mathcal{\ell}}$ $(\bigtriangleup \nu)$ before point source subtraction has a value around $10^{4}$ mK$^{2}$. This is seven order of magnitude stronger than the expected redshifted \HI\ signal. After subtracting the point-source model from the data, they have estimated the angular power spectrum of the residual data, which shows a power-law like feature corresponding to the DGSE.  They have found that the DGSE is  dominated at short baselines (large angular scales),  $U < 150$ and the clustering of point sources dominates in the range $150 < U < 2 \times 10^{3}$ and the residual point source Poisson fluctuations dominates at $U > 2 \times 10^{3}$. For a given angular multipole ($\mathcal{\ell}$), the foreground signal is expected to be smooth across frequency. However, \citet{Ali_2008} reported a oscillatory feature across frequency, which is most likely due to calibration error. 

\citet{paciga2011} put the first upper limit on the 21 cm power spectrum at redshift $z\sim 8.6$, using $\sim 40$ hours of observation with the legacy GMRT.  The chosen field was centered around a pulsar B0823+26, having a period of 0.53 seconds and an average flux of 350 mJy at 150 MHz. , whose on-pulse flux is about 6 Jy, brighter than all other sources in the field, making it a good calibrator. After calibration and RFI removal, they fit a piece-wise linear function to each baseline at each time stamp and subtract it to remove the contamination by bright spectrally smooth foregrounds. After subtracting the foregrounds, they cross-correlate multiple night's data set and estimate the power spectrum. The $2\sigma$ upper limit on \HI\ 21-cm power spectrum at redshift $z\sim8.6$ is (70 mK)$^{2}$ at $k=0.65$ $h$ Mpc$^{-1}$ \citep{paciga2011}. 

A revised upper limit using the same data was reported by \citet{paciga2013}. In the previous work, the authors did not account the signal loss associated with their foreground subtraction algorithm. In this work, they used singular value decomposition (SVD). After removing the first 4 SVD modes, which are mostly dominant by foregrounds, they quantified the signal loss by quantifying the transfer function between observed power after SVD modes subtraction and the fiducial 21-cm signal model. They put a revised $2\sigma$ upper limit about (248 mK)$^{2}$ at $k=0.5$ $h$ Mpc$^{-1}$ \citep{paciga2013}.

\begin{figure}[ht!]
    \centering
    \includegraphics[width=0.5\textwidth]{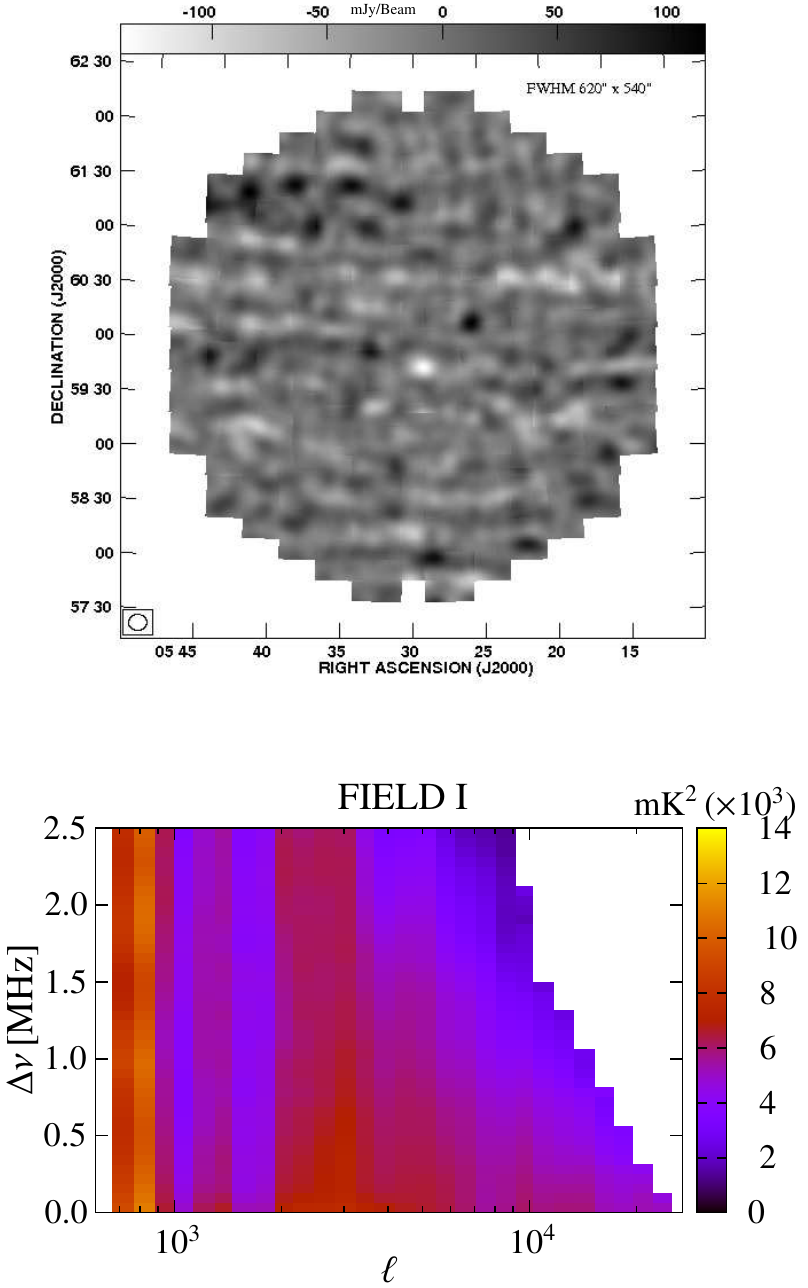}
        
    \caption{Top panel: The image of FIELD I ($\alpha_{2000}=5^{h}30^{m}00^{s} ,\delta_{2000}=60^{\circ}00'00''$ ) after subtracting the point sources and using a $uv$-taper at $|\u| = 200$. The synthesized beam width is $620 \times 540 ~\rm arcsec^{2}$. The diffuse structure is visible on $>10 ~\rm arcmin$ scale. Bottom panel: The measured $\cl$ $(\Delta \nu)$ of FIELD I. } 
    \label{fig:Abhik_2011}
\end{figure}

\begin{figure*}[ht!]
	\centering
	\includegraphics[width=0.9\textwidth]{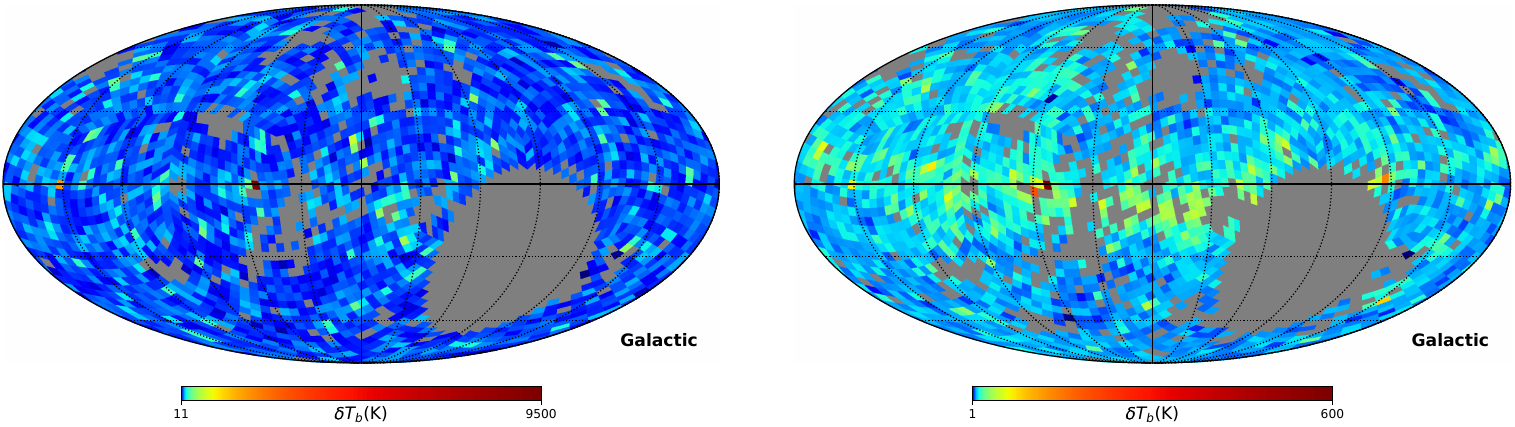}
	
	\caption{Shows the rms fluctuations of the brightness temperature ($\delta \rm T_{b}$) all over the sky at angular scale $l = 384$ ($0.47 \deg$). The left and right panels show the values of $\delta \rm T_{b}$ before and after subtracting the point sources from the data, respectively. } 
	\label{fig:TGSS_2020}
\end{figure*}

\cite{Ghosh_2012} have analyzed four different fields: FIELD I ($\alpha_{2000}=5^{h}30^{m}00^{s} ,\delta_{2000}=60^{\circ}00'00''$ ), FIELD II ($\alpha_{2000}=6^{h}00^{m}00^{s} ,\delta_{2000}=62^{\circ}12'58''$ ), FIELD III ($\alpha_{2000}=12^{h}36^{m}49^{s} ,\delta_{2000}=62^{\circ}12'58''$ ), FIELD IV ($\alpha_{2000}=1^{h}36^{m}48^{s} ,\delta_{2000}=41^{\circ}24'23''$ ) using GMRT observations at 150 MHz ($z\sim 8.4$). They  found that after subtracting the point sources from FIELD I data, the residual data is dominated by galactic synchrotron radiation at $\mathcal{\ell}$ range $ 253 \leqslant \l \leqslant 800 $ which they modelled as $\cl \propto \l^{-\beta}$. The estimated the fluctuations in the Galactic synchrotron emission is $\sqrt{\l (\l +1) \cl/(2 \pi)} \sim 10 \rm K$ at $\l = 800 $, corresponding to $\theta > 10'$. The residual point sources and artifacts dominates $\cl$ at smaller angular scales where  $\cl \sim 10^{3}$ mK$^{2}$ for $\l > 800$.  For other three fields, they found that the data is dominated by residual artifacts and they did not quantify the statistical properties of foregrounds for these fields. The point source subtracted residual map is shown in the top panel of Figure \ref{fig:Abhik_2011}.  \citet{Ghosh_2012} also estimated the the Euclidean normalized differential source counts as a function of flux density bins and found that the radio source population in the field is dominated by radio-loud AGNs and there is no signature of flattening in the source count at low flux density regime. They also quantified the the spectral variation of angular power spectrum of foregrounds  at a particular $\l$, $\cl$ $(\Delta \nu)$, using MAPS \citep{Kanan_2007}. The measured  $\cl(\Delta \nu)$ as a function of $l$ for FIELD I is shown in the bottom panel of Figure \ref{fig:Abhik_2011}. They found that $\cl$ $(\Delta \nu)$ shows a smooth frequency dependence over 2.5 MHz   $\Delta \nu$ range with $20-40$ percent variation. However, in addition to this they also found a oscillatory feature, similar to \citet{Ali_2008}, in the $\Delta \nu$ dependence of the measured $\cl$ $(\Delta \nu)$. In contrast to \citet{Ali_2008}, where point source model were subtracted above a flux cut off, \citet{Ghosh_2012} used a polynomial fit across frequency to model and subtract the foregrounds. However, if this oscillatory feature across frequency remained in the data due to calibration error or unknown systematics, then it will significantly affect any attempt to detect cosmological \HI\ signal.

\citet{Samir_2017} have analyzed two different pointings of TIFR all sky survey (TGSS; \citealt{Intema2017A&A...598A..78I} ) at  150 MHz near the Galactic plane ($9^{\circ} ,+10^{\circ}$) and ($15^{\circ} ,-11^{\circ}$) to characterize the statistical properties of DGSE. They  applied the novel 2-dimensional TGE to the low-frequency observation (150 MHz) with GMRT in order to quantify the angular power spectrum of the diffuse foreground from the measured visibility data set.  They  found that the measured angular power spectrum shows a power law behavior in the $\l$ range $ 240 \leqslant \l \leqslant 580$ and $ 240 \leqslant \l \leqslant 440 $  and the best fitted values of ($A,\beta $) are (356, 2.8) and (54, 2.2) for two different fields  respectively. Recently,  \citet{Samir_2020}  analyzed the whole TGSS all sky data to find out the fluctuations in  brightness temperature of galactic synchrotron radiation and to quantify the variation of amplitude of fluctuation at different galactic latitudes and longitudes. The rms fluctuation of the brightness temperature $\delta \rm T_{b}$ for the whole sky at angular scale $l = 384$ before and after the point source subtraction is shown in the Figure \ref{fig:TGSS_2020}. This will essentially help us to understand the relative contribution of synchrotron radiation at different parts of the sky. They found that $\cl$ decreases significantly as we move away from the galactic plane and $\cl$ is nearly symmetric in the Northern and Southern hemispheres except in the latitude range $15^{\circ} - 30^{\circ}$, which is the transition region from the disc dominated to the diffuse halo-dominated region. 

\begin{figure}[ht!]
	\begin{center}
		\includegraphics[width=0.48\textwidth]{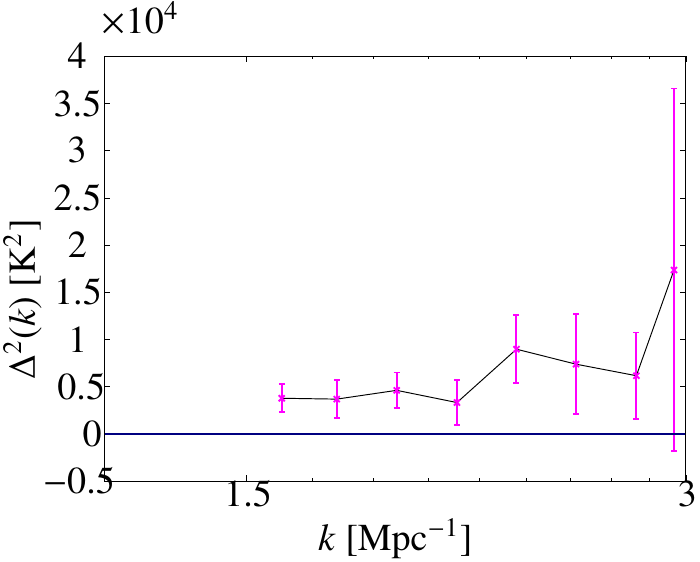}
		\caption{Shows the mean square brightness temperature fluctuations $\Delta^2(k)$ shown as a function of $k$ along with $2 \sigma$ error bars.}
		\label{fig:fin}
	\end{center}
\end{figure}

\citet{Pal21} have applied the MAPS based  TGE to the FIELD I data at 150 MHz ($z\sim 8.2$) of \citet{Ghosh_2012}. They first estimated the MAPS $\cl$ $(\Delta \nu)$ and from that they estimated the 21 cm power spectrum in $P (k_{\perp},k_{\parallel})$ domain. They obtained the $2\sigma$ upper limit on the dimensionless 21 cm power spectrum at $k=1.59$ Mpc$^{-1}$, $\Delta^{2}(k) = (72.66)^{2}$ K$^{2}$. The spherically averaged 3D power spectrum is shown in Figure \ref{fig:fin}. The estimated power spectrum is consistent with the expected noise, which shows that the analysis pipeline correctly estimates and subtracts the positive noise bias and giving a robust un-biased estimate of cosmological \HI\ 21 cm power spectrum. They also showed that using higher tapering of FoV, foreground contribution can be suppressed at large angular separation from the phase center.  

In addition to these observations corresponding to EoR redshift,  quite a significant amount of work have been done using GMRT at frequencies corresponding to the post-EoR epoch.  \citet{Ghosh_2011} used GMRT observation of the Hubble deep field North at 618 MHz to measure the fluctuations in the faint \HI\ 21 cm background. They, for the first time, put upper limit on [$ \bar{x}_{\mathrm{HI}} b_{\mathrm{H{\sc I}}}$] $\leq$ 2.9 ([$\Omega_{\mathrm{H{\sc I}}} b_{\mathrm{H{\sc I}}}$] $\leq$ 0.11) at $z \sim 1.32$, where $ \bar{x}_{\mathrm{HI}}$ is the mean neutral fraction. \citet{Arnab_GSB_2019a,Arnab_GWB_2019b} observed the ELAIS-N1 field using upgraded wideband correlator of  GMRT (uGMRT) at 300-500 MHz. They have characterized the spatial and spectral variation of DGSE and point-sources using this sensitive observation. \citet{Arnab_GSB_2019a} showed that how direction-dependent and -independent calibration affects the estimation of the angular power spectrum of DGSE in presence of  bright point sources far away from the phase center. After applying different calibration techniques to the data they showed  that using large tapering of the field of view (FoV),  TGE  can still reconstruct the DGSE power spectrum even in the presence of direction-dependent calibration errors. This study proves the robustness and effectiveness of TGE in presence of calibration systematic. For the first time, \citet{Arnab_GWB_2019b} have estimated the spectral characteristics of the angular power spectrum  of diffuse Galactic synchrotron emission (DGSE) over the wide frequency bandwidth  of $300-500$~MHz. The estimated spectral index of DGSE is consistent with previous all-sky monopole measurements,  however they also showed that there may be  a possible break in the DGSE spectrum at $\sim 400$ MHz.  But more sensitive observations using much wider bandwidth data is required to infer conclusively. Finally, \citet{Arnab_2021} divided the data in 8 MHz chunk and chose 4 such relatively less RFI dominated sub-bands to estimate the \HI\ power spectrum. They found that the spherically averaged power spectrum  is close to the theoretical thermal noise power, which shows that the estimated of \HI\  power is thermal noise limited.  The lowest limits, at $k \sim 1.0$ Mpc$^{-1}$, on spherically averaged \HI\ 21 cm power spectrum are (58.87 $\mathrm{mK})^{2}$, (61.49 $\mathrm{mK})^{2}$, (60.89 $\mathrm{mK})^{2}$, (105.85 $\mathrm{mK})^{2}$  at $z = 1.96,2.19,2.62$ and $3.58$, respectively. \citet{Arnab_2021} also constrained the product of neutral H{\sc i} mass density ($\Omega_{\mathrm{H{\sc I}}}$) and \HI\  bias ($b_{\mathrm{H{\sc I}}}$) to the underlying dark matter density field using the best limit on the \HI\  power spectrum. The  upper limits on [$\Omega_{\mathrm{H{\sc I}}} b_{\mathrm{H{\sc I}}}$] are 0.09,0.11,0.12,0.24 at $z=1.96,2.19,2.62,3.58$, respectively. 

\subsection{Calibration error and residual gain effects}

In an interferometric observation, the observed voltages hides the sky signal over the instrumental and line of sight effects. Various techniques with a known sky models are used \citep{1984ARA&A..22...97P, 1992ExA.....2..203W,2007ITSP...55.4497V, 2009ITSP...57.3512W} to estimate the modification in the signal and hence calibrate it. However, estimation of the instrumental and ionospheric effects, the gains, is subjected to the accuracy of the sky model and the sensitivity of the telescope. The residual gains are usually small and can be ignored. For the detection of the cosmological 21-cm signal, however, the residual gains can overwhelm the signal due to the presence of much larger foreground. Various calibration effects  for LOFAR-EoR, HERA and other  experiments are discussed in  \citet{2018MNRAS.478.1484G, 2016MNRAS.463.4317P, 2016JApA...37...35A, 2015MNRAS.453..925V, 2016MNRAS.458.3099V,2016RaSc...51..927M, 2010MNRAS.408.1029L,2021MNRAS.506.2066C, 2020MNRAS.499.5840D}. 
The effect of inaccurate models for sky based calibrations and its limitations are discussed in \citet{2016MNRAS.461.3135B, 2017MNRAS.470.1849E, 2019ApJ...875...70B}. Other instrumental effects like antenna position errors and variations in telescope beam with time and frequency on calibration solutions are discussed in \citet{2018AJ....156..285J, 2019MNRAS.487..537O, 2021MNRAS.506.2066C}.
Extragalactic point source contamination due to position errors in the sky model for bright sources, as well as the frequency-independent residual gain errors in interferometric calibration are studied in \citet{2009ApJ...703.1851D, 2010ApJ...724..526D}. 

A detailed investigation on the effect of time-correlated residual gain errors in presence of strong foreground is presented in \citet{2020MNRAS.495.3683K, 9113590} and \citet{2022MNRAS.512..186K}. Using simulated observations for GMRT Baseline configuration \citet{2020MNRAS.495.3683K} have shown that, even for a perfect foreground removal, the presence of time-correlated residual gain errors introduces a bias and  enhances  the uncertainty in the 21-cm power spectrum estimates. Further, with an analytical investigation of the problem, in presence of strong foregrounds and thermal noise \citet{2022MNRAS.512..186K} explore the importance to address this problem while designing the visibility based power spectrum estimators.

\begin{figure}[ht!]
    \centering
    \includegraphics[width=0.49\textwidth]{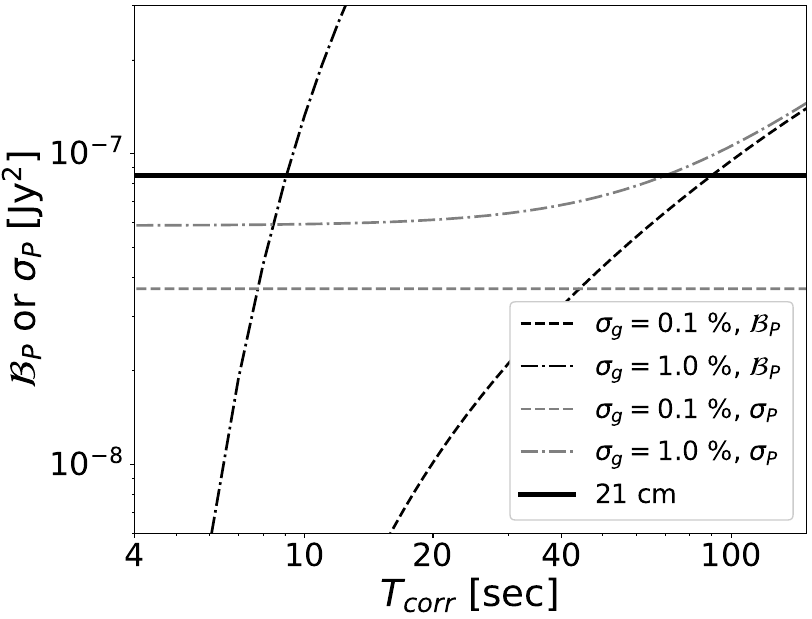}
    \caption{Shows the variation of $\mathcal{B}_P$ (thin black) and $\sigma_P$ (thin grey) against different correlation time with $\sigma_g = 0.1, 1.0$ \%. This considers a bandwidth of $16$ MHz, and an observation time of $400$ days. The plots are made at a baseline of $300$ $\lambda$. The thick black line corresponds to the expected EoR power spectrum \citep{Bharadwaj_2005}.} 
    \label{fig:Tcorr}
\end{figure}

The Figure~\ref{fig:Tcorr} show the bias $\mathcal{B}_P$ and uncertainty $\sigma_P$ in the power spectrum at a fixed baseline of  0.3 $k\lambda$ for 400 days observation at $150$ MHz for various levels of residual gains. The solid black line gives the expected 21-cm signal \citep{Bharadwaj_2005}. We see that for small values of gain errors $\sigma_g$, $\sigma_P$ is rather small and doesn't vary much with $T_{corr}$. For $\sigma_g = 1$ \% we see that both $\mathcal{B}_P$ and $\sigma_P$ increases with correlation time and $\mathcal{B}_P$ dominates of $\sigma_P$. This figure demonstrate that, in case, even if the error in the power spectrum is significantly low, the bias $\mathcal{B}_P$ can still be higher. Ignorance of the bias originated from the residual gains then  can confuse the scientific inference with a biased detection of the signal.

In a nutshell, given the advancements in design of estimators for the signal statistics, it is important to realize that the calibration issues are expected to provide further challenge in the detection of the 21-cm signal. 
Various hybrid calibration approaches, to minimize the effect of calibration errors, are also being developed and applied in different EoR experiments. \citet{2021MNRAS.503.2457B} have presented a hybrid calibration framework that unifies both the sky-based and redundant calibration and they show an improvement in calibration performance through simulations. A hybrid correlation calibration (CorrCal) scheme, to address the issue of sky-model errors and imperfect array redundancy, is also presented and applied on PAPER experiments data \citep{2022MNRAS.510.1680G}. A similar calibration approach is also discussed and applied on the MWA phase II data in \citet{2020PASA...37...45Z}. A precision bandpass calibration method namely CALAMITY is presented in \citet{2021arXiv211011994E}.
On the other hand, it seems also viable to design the statistical estimators of the 21-cm signal such that the calibration bias can be directly avoided. We are working on these aspects.

\section{Current upper limits and future prospects with SKA-Low}\label{sec:upper_lim}

\begin{figure*}[ht!]
	\centering
	\includegraphics[width=0.8\textwidth]{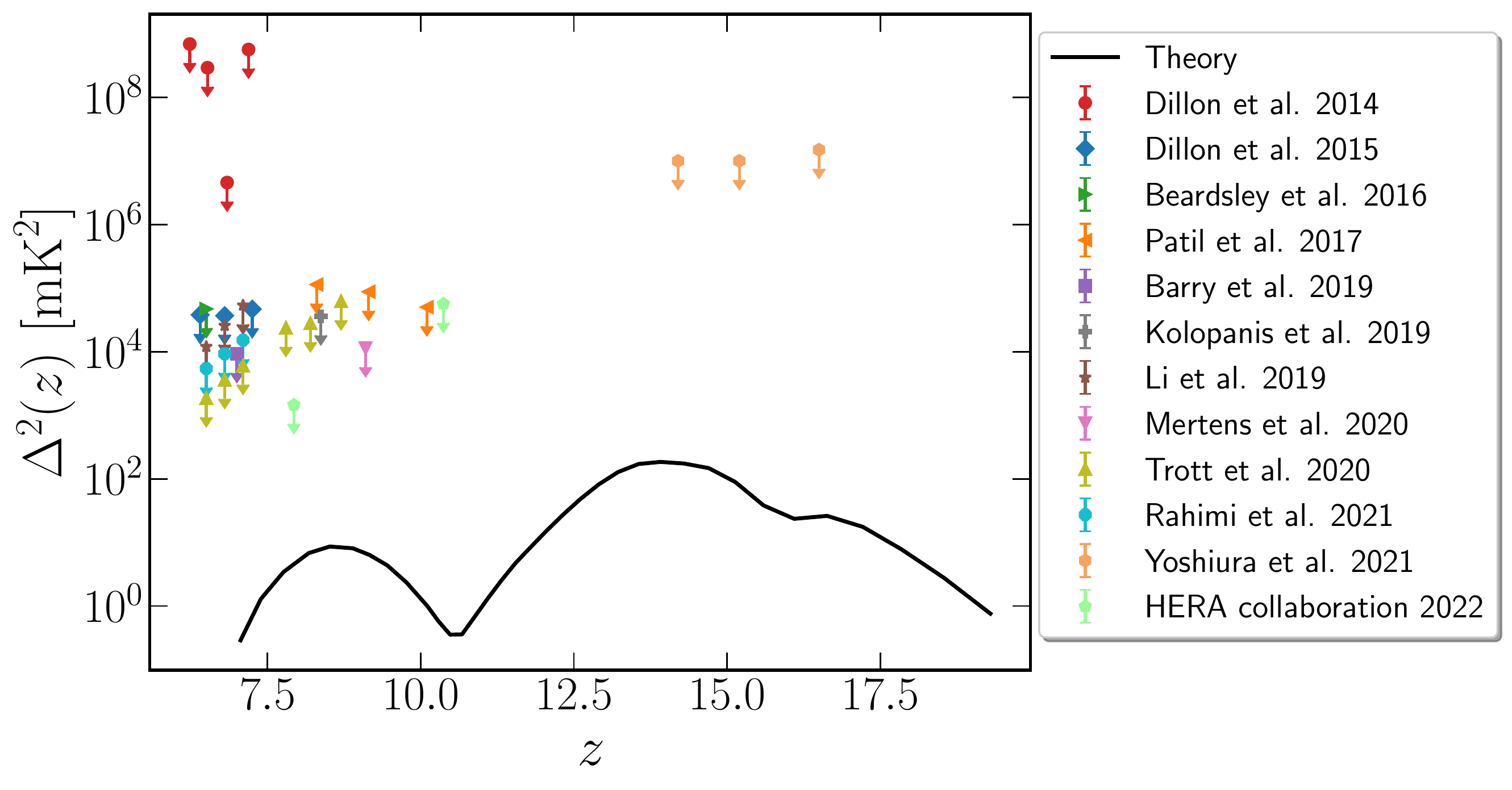}
	\caption{Shows a summery plot of the upper limits (in points) at $k\sim 0.1\,{\rm Mpc}^{-1}$, available to date, measured with the current instruments. The solid black line is a theoretical power spectrum estimated from a typical \textsc{GRIZZLY} simulation.}
	\label{fig:upper_lims}
\end{figure*}

There are several sensitive upper limits (see Figure \ref{fig:upper_lims}) placed on CD-EoR 21-cm power spectrum by current radio interferometers, such as MWA, LOFAR, HERA, using advanced data analysis algorithms and dedicated observation campaigns. The knowledge acquired with these observations will be used in future SKA-Low to detect the 21-cm power spectrum. Here we briefly discuss the current status of these telescopes.

{\bf LOFAR :-} The Low Frequency Array (LOFAR) \footnote{\url{https://www.astron.nl/telescopes/lofar/}} is a phased aperture array with tiles or stations spread over nine countries in Europe, centered in the Netherlands and operates in the range $30-240$ MHz \citep{vanhaarlem2013}. \citet{patil2017} placed the first upper limit on EoR using $13$ hours of data on North Celestial Pole (NCP) with LOFAR High Band Antenna (HBA) at redshift  range $z\sim 7.9-10.6$. The $2\sigma$ upper limit is $(79.6 {\rm mK})^{2}$ at $k=0.053$ $h$ cMpc$^{-1}$. \citet{gehlot2019} put the first upper limit on the power spectrum of \HI~21-cm brightness fluctuation from CD using $14$ hours of LOFAR Low Band Antenna (LBA). They used data on two different fields centered around NCP and 3C220.3 radio galaxy. The $2\sigma$ upper limits on \HI~power spectrum at redshift range $z\sim 19.8-25.2$ are $(14561 \,{\rm mK})^{2}$  and $(14886 \,{\rm mK})^{2}$ at $k=0.038$ $h$ cMpc$^{-1}$ for the 3C220 and NCP field respectively.  A more stringent and deep upper limit at $z\sim 9.1$ was placed by \citet{Mertens2020} using $141$ hours of data obtained with LOFAR HBA centered on NCP field. The best $2\sigma$ upper limit at $k=0.075$ $h$ cMpc$^{-1}$ is $(73 \,{\rm mK})^{2}$. However, all these results  have also shown excess power above theoretical thermal noise due to residual foreground emission, polarization leakage, chromatic calibration errors, ionosphere, or low-level radio-frequency interference etc \citep{Mertens2020,patil2017,gehlot2019, 2022MNRAS.509.3693M, 2022arXiv220302345G}. 

{\bf MWA :-} The Murchison Widefield Array (MWA)\footnote{\url{https://www.mwatelescope.org/}}, is also an aperture array, operational between $80-300$ MHz,  consisted of $128$ square ``tiles" of $4$\,m$\times4$\,m, distributed over $\sim3$ kms, which later got upgraded to $256$ tiles over $\sim 5$ kms \citep{tingay2013, Wayth_2018}. It is a precursor to the SKA-Low located in the Murchison Radio Observatory in Western Australia. \citet{Dillon_2015} set a upper limit of $3.7 \times 10^{4}$ mK$^{2}$ at $k=0.18$ $h$ Mpc$^{-1}$ around $z\sim6.8$ using $3$ hours of data with $128$-tile MWA. \citet{Ewall_MWA_2016} put upper limits on 21-cm brightness temperature fluctuation between the redshift range $z\sim 11.6 - 17.9$ using $3$ hours of observation with MWA-$128$ tiles. They achieved an upper limit of $10^{4}$ mK on comoving scales $k<0.5$ $h$ Mpc$^{-1}$. However, they found that their result is limited by calibration systematic \citep{Ewall_MWA_2016}. \citet{Beardsley_2016} have improved their analysis pipeline and put more stringent upper limit using $32$ hours of data at $z\sim 7.1$. They found that their estimated power spectrum is systematic limited and put a limit of $2.7 \times 10^{4}$ mK$^{2}$ at $k=0.27$ $h$ Mpc$^{-1}$. \citet{barry2019} used the same data set as \citet{Beardsley_2016} and presented much improved results, lowering the systematic effects by a factor of $2.8$ in power. Using newly developed analysis technique and RFI removal algorithm, they put a upper limit of $3.9 \times 10^{3}$ mK$^{2}$ at $k=0.2$ $h$ Mpc$^{-1}$ around $z\sim7$ using $21$ hours of MWA-$128$ data set. This improved the upper limit by almost an order of magnitude from their previous analysis. \citet{li2019} improved their calibration algorithm and RFI removal strategy and put the best upper limit of $2.39 \times 10^{3}$ mK$^{2}$ at $k=0.59$ $h$ Mpc$^{-1}$ around $z\sim 6.5$ using $40$ hours of  data set. \citet{Trott_2020} presented multi-redshift upper limits using four seasons of data obtained with MWA. Their best measurement yields $1.8 \times 10^{3}$ mK$^{2}$ at $k=0.14$ $h$ Mpc$^{-1}$ around $z\sim6.5$ using $110$ hours of data set. \citet{Yoshiura_2021} put an upper limit at higher redshifts, $z\sim 13-17$, using $15$ hours of MWA data set. The best upper limit they got about $6.6 \times 10^{6}$ mK$^{2}$ at $k=0.14$ $h$ Mpc$^{-1}$ around $z\sim15.2$.

\begin{figure*}
    \centering
    \includegraphics[width=0.9\textwidth,angle=0]{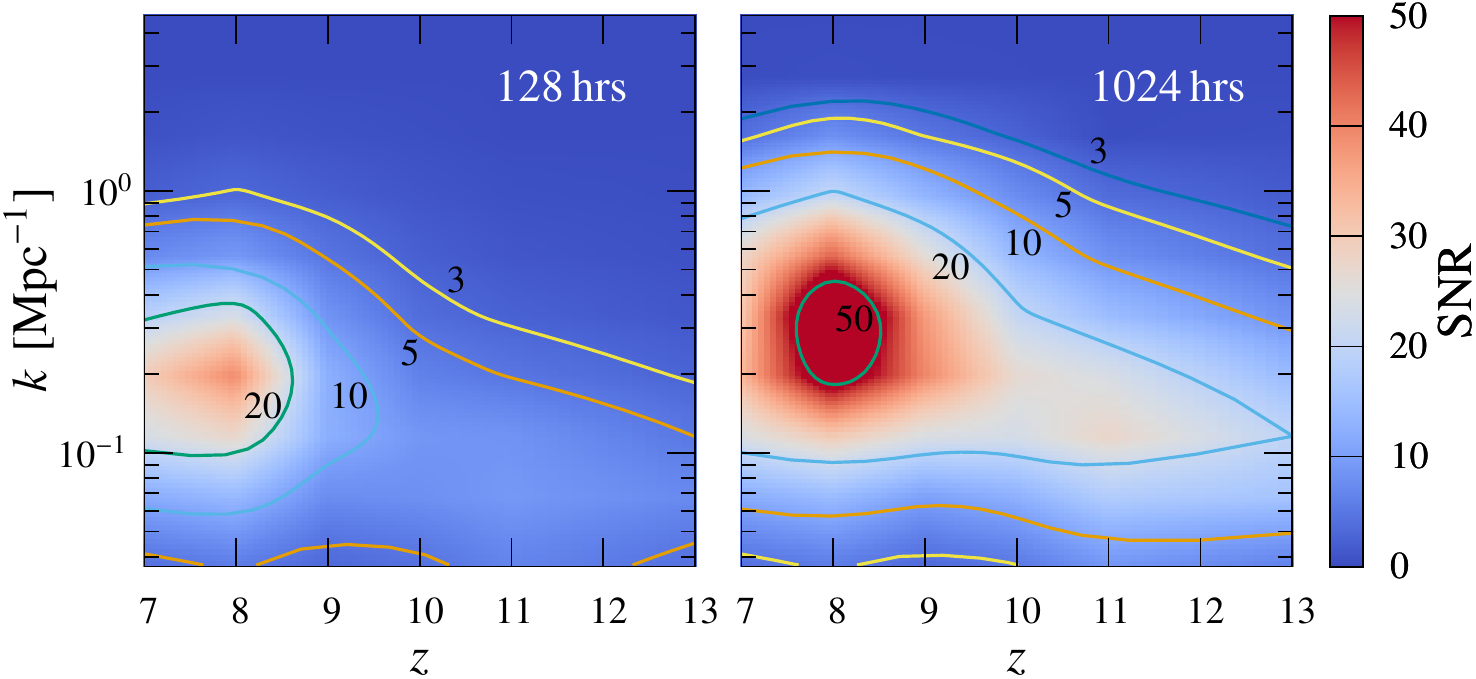}
    \caption{Shows the SNR predictions for measuring 3D EoR 21-cm power spectrum $\Delta^2(k,z)$ using $128$ hrs (left panel) and $1024$ hrs (right panel) of SKA-Low observations. This predictions includes the system noise contributions, however ignores the foreground effects. Credit: \citet{Shaw2019}}
    \label{fig:SKA_predic}
\end{figure*}

{\bf HERA :-}
The Hydrogen Epoch of Reionization Array (HERA) \footnote{\url{https://reionization.org/}} is an interferometric array of zenith pointing fixed dishes of diameter $14$ m located in Karoo desert, South Africa. The dishes are packed in a hexagonal configuration with nearly continuous core of $300$ m across.  HERA will operate in the range $50-250$\,MHz and will be built in a series of phases with simultaneous construction and observations. The full HERA  will consist of 350 dishes spread out to $\sim 850$ m,  once fully operational. Recently, \citet{HERA2021} reported results with $18$ nights of data ($\sim 36 \,{\rm hours}$) obtained with Phase-I of HERA, which commenced with $50$ antennas. They have put upper limits on the 21-cm power spectrum of $(30.76 \,{\rm mK})^{2}$ at $k=0.192$ $h$ Mpc$^{-1}$ at $z \sim 7.9$, and also  $(95.74 \,{\rm mK})^{2}$ at $k=0.256$ $h$ Mpc$^{-1}$ at $z \sim 10.4$ \citep{HERA2021}.

In terms of sensitivity and resolution SKA-Low is going to surpass its precursor telescopes which are mentioned earlier. It is planned to have $\sim 130,000$ log-periodic dipole antennas, distributed over $512$ aperture array stations each having $256$ dipole antennas. Construction of the dipoles have already been started in the Boolardy site in Western Australia. Roughly half of the stations will be located in the dense core within $1$ km diameter and the rest would be distributed on the three spiral arms around the central core, allowing the maximum baseline to extend up to $65$ kms. Each station will be $35$ m in diameter and the dipoles will operate within frequency bandwidth of $50-350$ MHz which covers the range of 21-cm signal from CD-EoR. Numerical predictions (see Figure \ref{fig:SKA_predic}) indicate that $\sim 100$ hours of SKA-Low observations will be sufficient to measure the EoR 21-cm power spectrum at $k\lesssim 0.3\,{\rm Mpc}^{-1}$ with $>5\sigma$. The SNR will increase rapidly for a longer observation ($\sim 1000$ hrs) thereby allowing $>5\sigma$ detection within a broader $k$ range. The results in Figure \ref{fig:SKA_predic} considers the system noise contribution assuming that foregrounds have been completely removed. The foreground avoidance decreases the overall SNR \citep[see e.g.,][]{Shaw2019}. A qualitatively similar prospect can be seen for the EoR 21-cm MAPS \citep{Mondal2020}. We refer the readers to these articles for details of the prospects with SKA-Low. Based on an older configuration of SKA-Low, \citet{koopmans2015} have shown that $\sim 1000$ hours of observations can detect the 21-cm power spectrum during mid and later stages of CD ($z\lesssim 15$) with SNR as high as $\sim 70$ at $k\sim 0.1~{\rm Mpc^{-1}}$. The SNR drops drastically towards the beginning of CD as the system noise increases as well as the signal decreases (see bottom panel of Figure \ref{fig:GRIZZLY_LC}) towards the larger redshifts.

\begin{figure*}
\centering
\includegraphics[width=0.9\textwidth, angle=0]{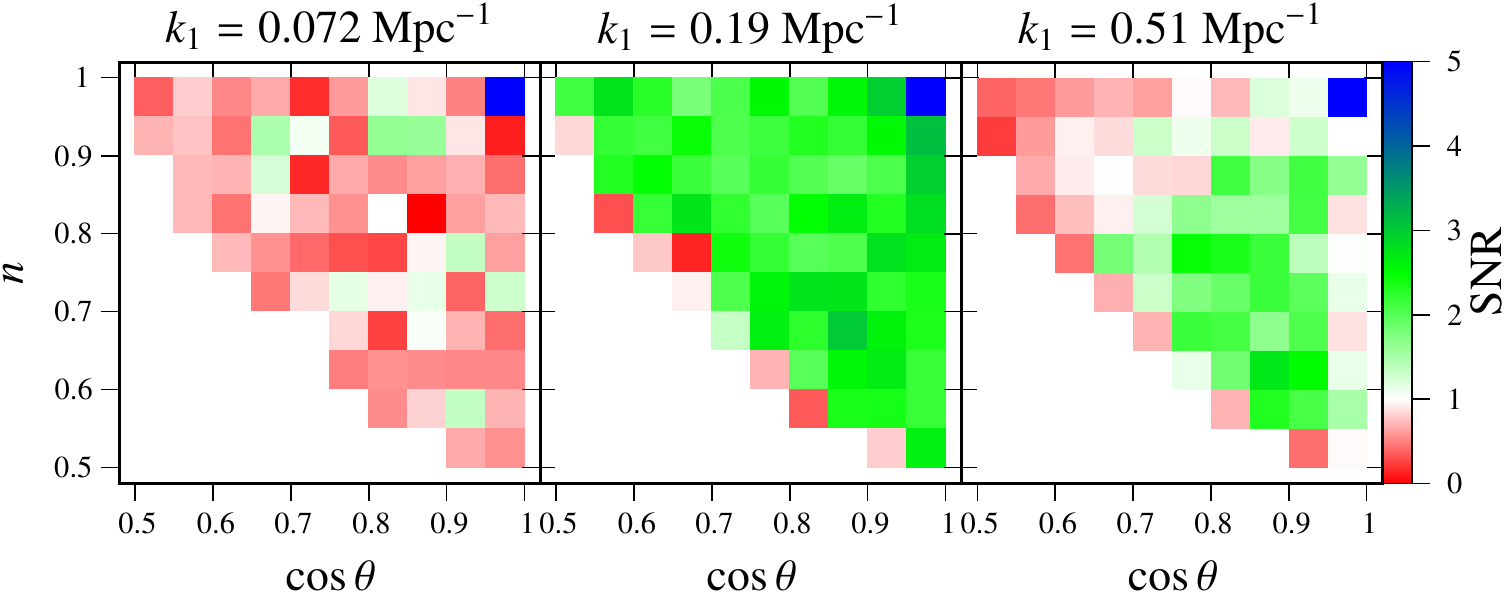}
\caption{Shows the SNR for detecting the reionization spherically-averaged bispectrum for all unique triangle configurations for three different triangle sizes. The predictions incorporate the system noise contribution of $t_{\rm obs}=1024$\,hrs of SKA-Low observations. Credit:\citet{mondal21}}
\label{fig:SNR_bispec}
\end{figure*}

Although bispectrum carries more information of the CD-EoR 21-cm signal as compared to the power spectrum, the measurement of bispectrum comparatively suffers more from the statistical errors \citep{watkinson21, mondal21}. This is evident from Figure \ref{fig:SNR_bispec} which shows the SNR predictions of EoR 21-cm bispectrum for all unique triangles of three different sizes ($k_1$), considering $1024$ hours of SKA-Low observations. A significant ($>5\sigma$) detection will is possible only for the squeezed limit $(n\rightarrow 1, \cos{\theta}\rightarrow 1)$ bispectrum where the signal value is itself larger. A less significant ($>3\sigma$) is expected to be possible for all unique triangles of intermediate sizes ($k_1\sim 0.1~{\rm Mpc}^{-1}$). However, for the SNR drops for both the smaller ($k_1\lesssim 0.1~{\rm Mpc}^{-1}$) and ($k_1\gtrsim 1~{\rm Mpc}^{-1}$), respectively due to prevailing cosmic variance and the system noise. Hence, either a high-resolution or a longer observation or a combination of both will be needed for a reliable measurement of the CD-EoR 21-cm bispectrum from the observations.


\section{Summary and Future scope}\label{sec:summ}

The most successful theory of structure formation states that the universe underwent changes in its thermal and ionized phases during cosmic evolution. The adiabatic expansion of the universe made it cold and neutral from a hot dense state at $z\approx 1100$ for the first time. Afterwards, the matter perturbations grew under self-gravity and eventually collapsed to form the first structures in the universe during the cosmic dawn (CD). The X-rays from the first stars, quasars and galaxies propagated out and heated the IGM during CD. This is followed immediately by the epoch of reionization (EoR) during which the UV photons from the first sources travelled into the IGM and ionized the cold atomic hydrogen (\HI). CD and EoR together marks the last transition in the thermal and ionization states of the universe. The CD-EoR is also enriched in information of the first sources which evolved into the structures that we see today. Studying these epochs will help us to bridge the gap in our understanding of the structure formation history. In this review, we discuss the importance of these epochs and understand the physical processes and the sources which drive the phase change. We also discuss the observational probes of CD-EoR, challenges and remedies.

Our current understanding of CD-EoR is only limited by few indirect observations which provide some loose constraints over the beginning and the end of these epochs. A dedicated direct observation is mandatory to answer various fundamental questions regarding these epochs such as the exact timing of these epochs, the properties of the sources involved and their evolution, how did reionization and heating progressed and changed the topology of the IGM, etc. The redshifted 21-cm radiation emitted due to hyperfine transition in the ground state of \HI, acts as the most promising direct probe to CD-EoR. We aim to observe the fluctuations in 21-cm brightness temperature and its evolution as CD-EoR signal. The first sources leave their imprint on the CD-EoR 21-cm signal on cosmological scales by heating and ionizing the IGM. Several current telescopes such as uGMRT, LOFAR, NenuFAR, MWA and HERA as well as the future SKA-Low aim to map this cosmological signal from CD-EoR. However the CD-EoR 21-cm signal is inherently $3-4$ orders of magnitude weaker as compared to the contamination from the galactic and extra-galactic foregrounds. A state-of-art methodology and precise knowledge of foreground are required to separate out the signal. Also, the system noise, ionospheric turbulence and other instrumental systematics adds up to the foreground contamination making direct mapping of the CD-EoR 21-cm signal very difficult for the current instruments we have. Therefore, the primary goal of the current CD-EoR experiments, including first phase of SKA-Low, is to measure the signal using statistical estimators such as variance, power spectrum, bispectrum, etc.

A beforehand knowledge of the CD-EoR 21-cm signal and its hindrances are crucial for interpreting the observed data. There have been several theoretical studies which use analytical or numerical prescriptions to simulate and understand the nature of the signal and its dependence on the different source properties. The state-of-the-art radiative transfer simulations model the signal accurately by incorporating the detailed physics during CD-EoR, but they are computationally expensive. The approximate but faster semi-numerical simulations provide a helpful alternative for exploring vast range of CD-EoR models. These different models are based on our current understanding of CD-EoR guided by the indirect observations. Several previous works have used simulations to make sensitivity predictions to measure the CD-EoR statistics in the context of present and future telescopes. These prediction studies are helpful in strategizing future observations.

Although the properties of cosmological 21-cm signal are believed to be well understood through the state-of-the-art large-scale simulations, however, the astrophysical foregrounds are poorly constrained. The major challenge for ongoing experiments and upcoming SKA-Low is separating the cosmological signal from foregrounds, which are $10^{3}-10^{5}$ times larger. Almost all foreground separation techniques use the fact that foregrounds are smooth as a function of frequency, whereas cosmological 21-cm signal decorrelates rapidly as each frequency corresponds to a different radial slice of the universe. However, real-world  problems, such as gain calibration errors, ionospheric fluctuations, variation of beam, instrumental systematics, etc. makes it difficult to separate smooth foreground components from the cosmological signal of interest. Hence it is imperative to characterize foregrounds and understand systematics through deep low-frequency observations with SKA pathfinder telescopes, such as GMRT.  

In addition, it is essential to develop a robust estimator which will give an unbiased estimate of the power spectrum of the signal. We have developed different visibility-based estimators over the past decade, which can address most of the real-world challenges, such as ($1$) reducing the noise bias by correlating the visibilities in nearby baselines, ($2$) reducing the computation cost by gridding the visibilities, ($3$) reducing the effect of the compact sources outside the field-of-view by tapering the response of antenna, ($4$) a novel algorithm to subtract effect of resolved compact sources in the field-of-view by implementing complex tapering function in the image plane and ($5$) using a MAPS-based algorithm to mitigate effect of flagged frequency channels. All of these estimators are tested against simulated observations and have been used to estimate the power spectrum of compact sources and galactic synchrotron radiation at $\lesssim 600~ {\rm MHz}$ frequencies.

Even if we are able to remove the foregrounds perfectly from the data set, the gain calibration error can still pose a hindrance to detect the cosmological signal. It is extremely challenging to calibrate the instrumental and ionospheric effects at low-frequencies, the precision of which depends on the sky model and instrument sensitivity. Since the dynamic range required to detect the cosmological signal is huge, \textit{i.e.}  $\sim 10^{5}$, any residual gain errors can overwhelm the 21-cm signal. We have shown that time correlated gain error, at the level of a few percent, can give rise to a bias and enhances the uncertainty in the 21-cm power spectrum estimates. This signifies the fact that to have a statistically robust detection of the cosmological \HI~signal, we should be able to precisely calibrate all instrumental effects.  

The SKA-Low is going to be a giant leap in terms of sensitivity as compared to the current radio interferometers and its precursors. 
With its improved sensitivity, SKA-Low is expected to push the field of CD-EoR cosmology towards a new edge in future. Also a parallel development of efficient algorithms/estimators are required to extract the CD-EoR signal statistics from the contaminated data. One can also do a cross-correlation study with other line emissions such as Ly-$\alpha$, ${\rm H}\alpha$, CO and CII, etc. This can provide additional constraints over the different reionization models.




\section*{Acknowledgements}
We thank Rajesh Mondal, Somnath Bharadwaj, Kanan K. Datta and Suman Majumdar for fruitful discussions and suggestions in shaping this article. We would like to thank all the collaborators who are part of the projects presented in this article. We would also like to acknowledge the support and resources from the following facilities used for parts of the works presented here: Param Shivay and Param Shakti facilities as a part of the National Super-computing Mission, Computing facilities at the IIT Kharagpur, IIT Indore and IIT (BHU), Varanasi, the upgraded Giant Meterwave Radio Telescope facility etc. RG and AKS acknowledge financial support by the Israel Science Foundation (grant no. 255/18). JK would like to acknowledge the University Grant Commission, Government of India for financial support. AG would like to acknowledge IUCAA, Pune for providing support through the associateship programme.

\vspace{-1em}


\bibliography{refs}

\end{document}